\documentclass[12pt, preprint]{emulateapj}
\usepackage{graphicx}
\usepackage{epstopdf}
\newcommand{\asec}      {\mbox{$^{\prime \prime}  $} }

\begin{document}
\title{Spectroscopy of luminous $z>7$ galaxy candidates and sources of contamination in $z>7$ galaxy searches$^{\star}$}
 \author{P. Capak\altaffilmark{1,2}, 
 B. Mobasher \altaffilmark{3},
 N.Z. Scoville\altaffilmark{2},
 H. McCracken\altaffilmark{4}, 
 O. Ilbert \altaffilmark{5},
 M. Salvato\altaffilmark{2}, 
 K. Men\'{e}ndez-Delmestre \altaffilmark{6},
 H. Aussel \altaffilmark{7},
 C. Carilli \altaffilmark{8},
 F. Civano \altaffilmark{9},
 M. Elvis \altaffilmark{9},
 M. Giavalisco \altaffilmark{10},
  E. Jullo\altaffilmark{1},
 J. Kartaltepe \altaffilmark{11},
  A. Leauthaud \altaffilmark{12},
 A.M. Koekemoer \altaffilmark{13},
  J.-P. Kneib \altaffilmark{14},
  E. LeFloch\altaffilmark{7},
 D.B. Sanders\altaffilmark{11},
 E. Schinnerer\altaffilmark{15},
 Y. Shioya\altaffilmark{16},
 P. Shopbell\altaffilmark{2},
 Y.Tanaguchi\altaffilmark{16},
 D. Thompson\altaffilmark{17},
 C.J. Willott\altaffilmark{18}
}

\altaffiltext{1}{Spitzer Science Center, 314-6 Caltech, 1201 E. California Blvd. Pasadena, CA, 91125}
\altaffiltext{2}{Department of Astronomy, 249-17 Caltech, 1201 E. California Blvd., Pasadena, CA, 91125}
\altaffiltext{3}{Department of Physics and Astronomy, University of California, Riverside, CA, 92521, USA}
\altaffiltext{4}{Institut d'Astrophysique de Paris, UMR7095 CNRS, Universit\`{e} Pierre et Marie Curie, 98 bis Boulevard Arago, 75014 Paris, France}
\altaffiltext{5}{Laboratoire d'Astrophysique de Marseille, BP 8, Traverse du Siphon, F-13376 Marseille Cedex 12, France}
\altaffiltext{6}{Carnegie Observatories, 813 Santa Barbara St., Pasadena, CA, 91101, USA}
\altaffiltext{7}{AIM Ð UnitŽ Mixte de Recherche CEA Ð CNRS Ð UniversitŽ Paris VII Ð UMR n¡ 7158}
\altaffiltext{8}{National Radio Astronomy Observatory, P.O. Box 0, Socorro, NM 87801-0387, USA}
\altaffiltext{9}{Harvard-Smithsonian Center for Astrophysics, 60 Garden St., Cambridge, MA, 02138, USA}
\altaffiltext{10}{Astronomy Department, University of Massachusetts, Amherst, MA 01003}
\altaffiltext{11}{Institute for Astronomy, University of Hawaii, 2680 Woodlawn Drive, Honolulu, HI, 96822, USA}
\altaffiltext{12}{LBNL \& Berkeley Center for Cosmological Physics, University of California, Berkeley, CA 94720, USA}
\altaffiltext{13}{Space Telescope Science Institute, 3700 San Martin Drive, Baltimore, MD 21218, USA}
\altaffiltext{14}{Laboratoire d'Astrophysique de Marseille, CNRS-Universit\`{e} Aix-Marseille, 38 rue F. Joliot-Curie, 13388 Marseille Cedex 13 France}

\altaffiltext{15}{Max-Planck-Institut fŸr Astronomie, Kšnigstuhl 17, D-69117 Heidelberg, Germany}
\altaffiltext{16}{Research Center for Space and Cosmic Evolution, Ehime University, 2-5 Bunkyo-cho, Matsuyama 790-8577, Japan}
\altaffiltext{17}{Large Binocular Telescope Observatory, University of Arizona, 933 N. Cherry Ave., Tucson, AZ, 85721, USA}
\altaffiltext{18}{Herzberg Institute of Astrophysics, National Research Council, 5071 West Saanich Rd., Victoria, BC V9E 2E7, Canada}
\altaffiltext{$\star$}{Based on observations with:  the W.M. Keck Observatory, which is operated as a scientific partnership among the California Institute of Technology, the University of California and the National Aeronautics and Space Administration and made possible by the generous financial support of the W.M. Keck Foundation; the Spitzer Space TelescopÄcoe, which is operated by the Jet Propulsion Laboratory, California Institute of Technology under a contract with NASA; the Canada-France-Hawaii Telescope with WIRCam, a joint project of CFHT, Taiwan, Korea, Canada, France, at the Canada-France-Hawaii Telescope (CFHT) which is operated by the National Research Council (NRC) of Canada, the Institute National des Sciences de l'Univers of the Centre National de la Recherche Scientifique of France, and the University of Hawaii; the United Kingdom Infrared Telescope operated by the Joint Astronomy Centre on behalf of the Science and Technology Facilities Council of the U.K;  the Subaru Telescope, which is operated by the National Astronomical Observatory of Japan;  the Canada-France-Hawaii Telescope with MegaPrime/MegaCam operated as a joint project by the CFHT Corporation, CEA/DAPNIA, the National Research Council of Canada, the Canadian Astronomy Data Centre, the Centre National de la Recherche Scientifique de France, TERAPIX and the University of Hawaii; the NASA/ESA {\em Hubble Space Telescope}, obtained at the Space Telescope Science Institute, which is operated by AURA Inc, under NASA contract NAS5-26555; the XMM-Newton, an ESA science mission with instruments and contributions directly funded by ESA Member States and NASA;  the Chandra X-ray observatory, which is operated by the Smithsonian Astrophysical Observatory for and on behalf of the National Aeronautics Space Administration under contract NAS8-03060; the National Radio Astronomy Observatory which is a facility of the National Science Foundation operated under cooperative agreement by Associated Universities, Inc ; the 30 m telescope of the Institute for Radio- astronomy at Millimeter Wavelengths ( IRAM ), which is funded by the German Max-Planck-Society, the French CNRS, and the Spanish National Geographical Institute.}


\begin{abstract}
We present three bright $z^+$ dropout candidates selected from deep Near-Infrared (NIR) imaging of the COSMOS 2 square degree field.   All three objects match the $0.8-8\mu$m colors of other published $z>7$ candidates but are three magnitudes brighter, facilitating further study.  Deep spectroscopy of two of the candidates covering $0.64-1.02\mu$m with Keck-DEIMOS and all three covering 0.94-1.10$\mu$m and 1.52-1.80$\mu$m with Keck-NIRSPEC detects weak spectral features tentatively identified as Ly-$\alpha$ at $z=6.95$ and $z=7.69$ in two of the objects.  The third object is placed at $z\sim1.6$ based on a 24$\mu$m and weak optical detection.  A comparison with the spectral energy distributions of known $z<7$ galaxies, including objects with strong spectral lines, large extinction, and large systematic uncertainties in the photometry yields no objects with similar colors.  However, the $\lambda>1\mu$m properties of all three objects can be matched to optically detected sources with photometric redshifts at $z\sim1.8$, so the non-detection in the $i^+$ and $z^+$ bands are the primary factors which favors a $z>7$ solution.  If any of these objects are at $z\sim7$ the bright end of the luminosity function is significantly higher at $z>7$ than suggested by previous studies, but consistent within the statistical uncertainty and the dark matter halo distribution.  If these objects are at low redshift, the Lyman-Break selection must be contaminated by a previously unknown population of low redshift objects with very strong breaks in their broad band spectral energy distributions and blue NIR colors.  The implications of this result on luminosity function evolution at high redshift is discussed.  We show that the primary limitation of $z>7$ galaxy searches with broad filters is the depth of the available optical data.
\end{abstract}

\section{Introduction }\label{s:introduction}
The $z>7$ universe holds the keys to many open questions about galaxy formation and evolution.  Cold Dark Matter (CDM) models predict that galaxies assemble through merging of dark matter haloes \citep{2005Natur.435..629S,2000MNRAS.319..168C,1993MNRAS.264..201K} and the distribution of initial halo masses is strongly constrained by the Cosmic Microwave Background (CMB) measurements \citep{2003ApJS..148....1B}.  However, populating the dark matter halos with galaxies is done in a semi-analytic manner with many free parameters that are tuned to observations.  At $z\simeq7$ the universe was less than a billion years old, leaving little time for galaxies to evolve, so the luminosities and masses of galaxies at $z>7$ are a direct probe of the assumptions underlying galaxy formation scenarios in structure formation models. 

Advances in red-sensitive Charge-Coupled-Devices (CCDs) have allowed studies of the universe out to $z\sim7$ \citep[e.g.][]{2009ApJ...706.1136O,2008ApJ...677...12O,2008ASPC..399...61I}; however, accessing higher redshifts is difficult due to the lack of wide field near-infrared cameras and high sky backgrounds beyond 0.8$\mu$m in the case of ground based telescopes.  Ultra-deep studies from space with Near Infrared Camera and Multi-Object Spectrometer (NICMOS) and Advanced Camera for Surveys (ACS) on the Hubble Space Telescope (HST) have constrained  the faint-end of the luminosity function  \citep{2008ApJ...686..230B,2010ApJ...709L..16O} with deep ground based studies and NICMOS parallel surveys placing useful limits on its bright end \citep{2008MNRAS.386..370S,Mannucci:2007p237,Henry:2009p256},  and lensing surveys also providing useful constraints on fainter galaxies  \citep{2004ApJ...607..697K, 2008ApJ...685..705R,2008ApJ...678..647B, 2009ApJ...690.1764B}  albeit with significant uncertainty due to the small areas covered.  Unfortunately, the selection function is complicated due to contamination from faint galaxies at $1.4<z<2.5$ where the galaxy population is also poorly understood and the fact that no objects have been spectroscopically confirmed \citep{2008MNRAS.386..370S}. 

	In this paper we use deep NIR imaging of the 2 square degree Cosmic Evolution Survey or COSMOS field with the Wide-Field Infrared Camera (WFCAM) instrument on United Kingdom Infrared Telescope (UKIRT) and the Wide-Field Infrared Camera (WIRCAM) instrument on the Canada-France-Hawaii-Telescope (CFHT) along with deep Spitzer Infrared Array Camera  (IRAC) and and Multi-Band Imaging Photometer (MIPS) imaging to color select candidate z-dropout galaxies.  We find three candidates brighter than $J_{AB}<23.5$, which have 0.9-8$\mu$m SEDs consistent with fainter $z>7$ candidates discovered in the GOODS and UDF fields \citep{2008ApJ...686..230B,2008MNRAS.386..370S,2010ApJ...709L..16O} and have mid-infrared (MIR) colors consistent with the selection proposed in \citet{2008MNRAS.386..370S}.  We then present deep spectra of these objects obtained with Keck-DEIMOS in the $0.64-1.02\mu$m  range and Keck-NIRSPEC in the 0.94-1.10$\mu$m and 1.52-1.80$\mu$m range.  In addition to these three candidates we find over 3500 objects which are at $z\sim1.8$ and would be selected as $z>7$ objects were it not for the large difference in depth between the near-infrared and optical bands.  Using this sample of interlopers, we discuss the sources of contamination for $z>7$ studies and strategies for removing them.
	
	Throughout this paper we use AB magnitudes or fluxes in units of Jansky and a $\Omega_m=0.3$, $\Omega_v=0.7$, H$_o=70$ km s$^{-1}$ Mpc$^{-1}$ cosmology.

\begin{figure}
\includegraphics[scale=0.34]{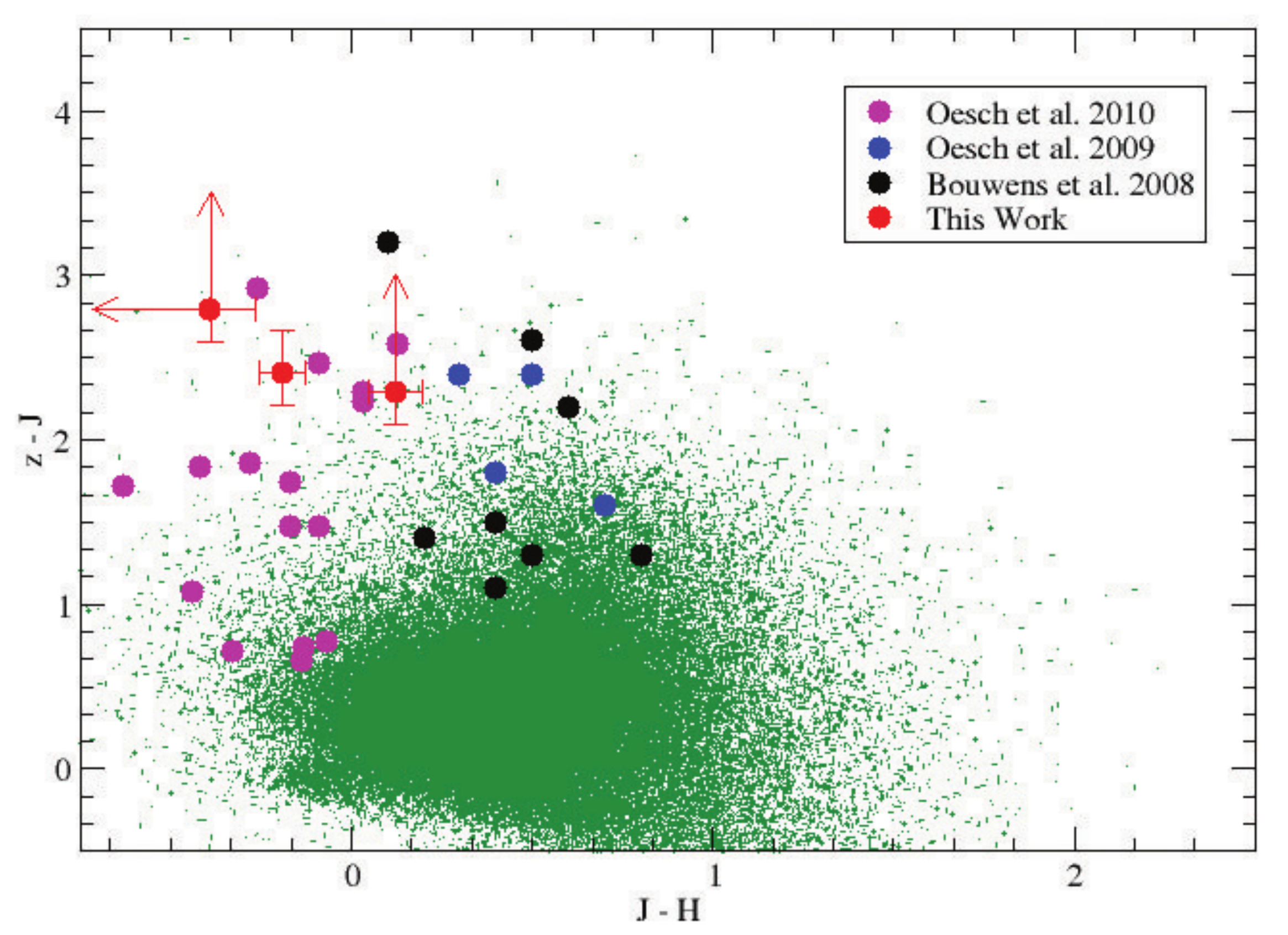} \\
\vskip 0.2in
\includegraphics[scale=0.34 ]{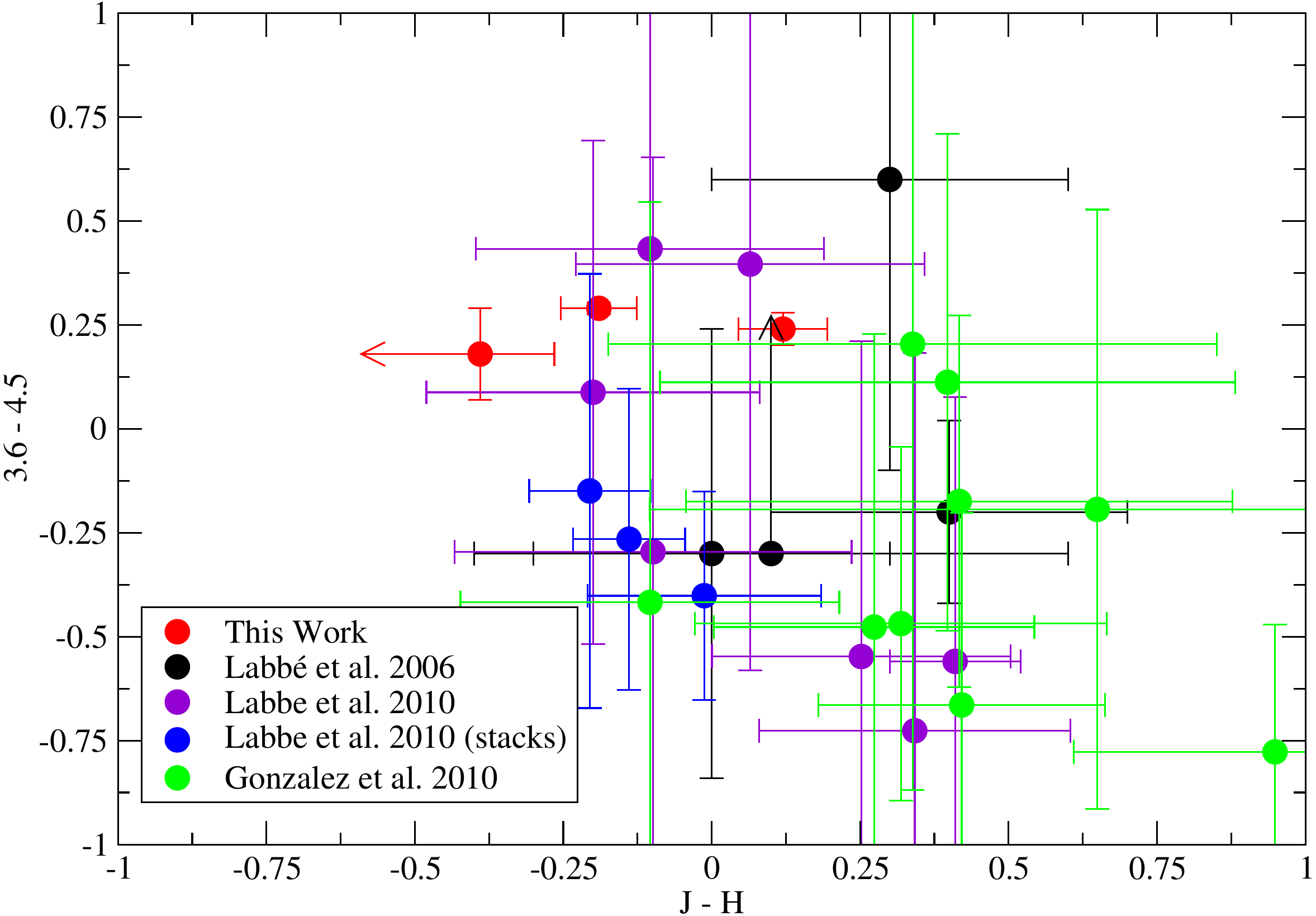}
\caption{\label{f:colors}The infrared colors of our $z>7$ candidates are shown with objects from the literature and foreground objects. In the top pannel the $J-H$ and $z^+-J$ colors of our candidates (red points) along with \citet{2008ApJ...686..230B} (black points), \citet{2009ApJ...690.1350O}( blue points)  and WFC-UDF \citet{2010ApJ...709L..16O} (magenta points) are shown.  The redder colors in \citet{2008ApJ...686..230B} and  \citet{2009ApJ...690.1350O} are due to the use of F110W, which is considerably wider and bluer than the J  and F125W band used in this work and the WFC3-UDF respectively.  The colors of optically detected sources at $z<6$ with $>10\sigma$ detections in $J$, $H$, and $K_s$ are shown as small green points.  In the bottom panel $J-H$ and $3.6-4.5$ colors of our candidates (red points) and those from \citet{2010ApJ...708L..26L} (blue and purple points), \citet{2010ApJ...713..115G} (Green Points), and \citet{2006ApJ...649L..67L} (black points) are shown.  The objects presented here have colors consistent with those selected in deeper surveys.  }
\end{figure}

\begin{deluxetable}{ccccc}
\tabletypesize{\scriptsize}
\tablecaption{Positions\label{t:objpos}}
\tablehead{
\colhead{ID} & \colhead{RA(J2000)} & \colhead{DEC(J2000)} & \colhead{FWHM \asec} & \colhead{Stellarity}
}
\startdata
1	&	$10^h 01^m 36.317^s$	&	+1:37:00.42 & 2.5 & 0.98\\
2	&	$9^h 59^m 25.507^s$	&	+2:21:27.79 & 1.5 & 0.96\\
3	&	$9^h 59^m 14.609^s$	&	+2:27:36.57 & 2.0 & 0.53\\
\enddata
\end{deluxetable}

\begin{figure*}
\begin{center}
\includegraphics[scale=0.8]{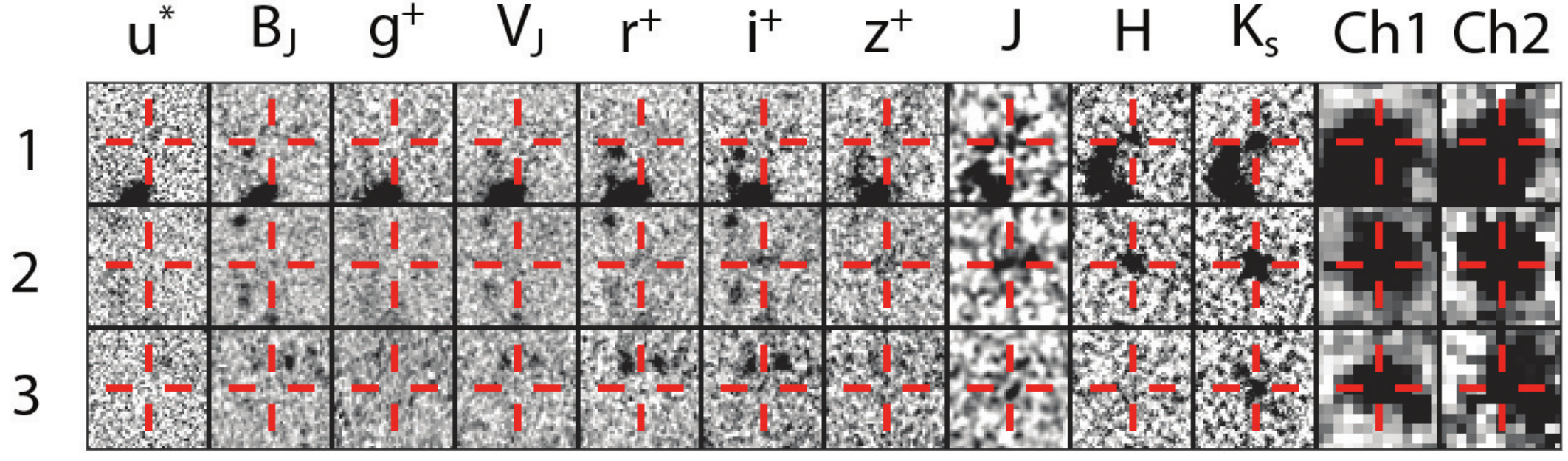} 
\caption{Postage stamps of the three candidate objects covering $7.5^{\prime\prime}\times7.5^{\prime\prime}$  in $u^*$, $B_J$, $g^+$, $V_J$, $r^+$, $i^+$, $z^+$, $J$, $H$, $K_s$, IRAC Ch1 (3.6$\mu$m), and IRAC Ch2 (4.5$\mu$m) bands scaled to a common noise level and S/N stretch in all images.  Notice the absence of the objects blueward of $J$ band despite the significant increase in the image depth.\label{stamps}}
\end{center}
\end{figure*}

\section{Imaging Data and Object Selection}

	The 2 square degree COSMOS field is covered by a combination of 30 broad, intermediate, and narrow Ultraviolet (UV), Optical, and NIR filters including data from Galex, CFHT, Subaru, HST and Spitzer.  These observations are described in detail in \citet{2007ApJS..172..468Z} (GALEX UV), \citet{capak-data,capak-smg} (Subaru, CFHT, and UKIRT optical and NIR), \citet{2007ApJS..172..196K} and \citet{scoville-hst} (HST optical and NIR), \citet{sanders-spitzer} (Spitzer near and mid-Infrared).  In addition there are extensive, X-ray data obtained with the Chandra X-ray observatory \citep{2009ApJS..184..158E} and XMM \citep{2007ApJS..172...29H}, sub-mm data from the IRAM 30m \citep{2007ApJS..172..132B} , along with 20cm  \citep{eva-vla} and 90cm (Smolcic et al. in prep) radio images from the Very Large Array (VLA).  	
	
	Throughout this paper we follow the filter notation used in \citet{capak-data} .  In particular Subaru-Sloan filters are noted with the super-script $^+$ (e.g. $z^+$)  and CFHT Megaprime filters are denoted with the super-script $^*$ (e.g. $u^*$) to differentiate them from each other and the true Sloan Digital Sky Survey (SDSS) filters which have somewhat different bandpasses.
		
	This paper also makes use of new deep $J$ band data obtained with the WFCAM instrument on UKIRT \citep{Hirst:2006p751}, along with $H$ and $K_s$ data obtained with WIRCAM on CFHT \citep{Puget:2004p698} telescopes as part of the COSMOS survey \citep{McCrakenK,capak2010data}.  The $J$, $H$ and $K_s$ data were taken over the course of three years with over 500 exposures at each position.  This is more than sufficient to remove cosmic rays, moving objects, satellites, and other transient objects that may be mistaken for high-z objects. Electronic cross-talk artifacts near bright objects are a significant problem in all three data sets.  The nature of the cross-talk in the $H$ and $K_s$ data allow the artifacts to be modeled and removed in all but the most severe cases \citep{McCrakenK}.  Any remaining cross talk is easily identified because it consists of 64 equally bright/dark images at 64 detector pixel intervals in Declination.  Cross-talk artifacts are still present in the UKIRT WFCAM data.  In WFCAM data the cross-talk artifacts occur at 128 detector pixel intervals in either Right Ascension or Declination from bright objects depending on position within the array and can be identified by their distinct positive-negative signature.

\begin{figure}
\begin{center}
\includegraphics[scale=0.8]{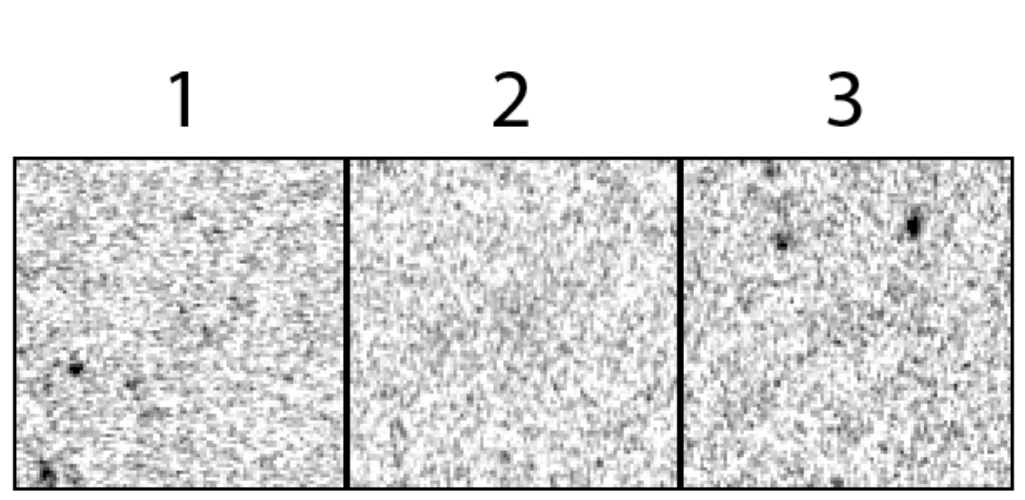}
\caption{ACS F814W postage stamps of the three candidate objects covering $5^{\prime\prime}\times 5^{\prime\prime}$.  Note the lack of any significant detection \label{acs-stamps}}
\end{center}
\end{figure}

\begin{figure}
\begin{center}
\includegraphics[scale=0.8]{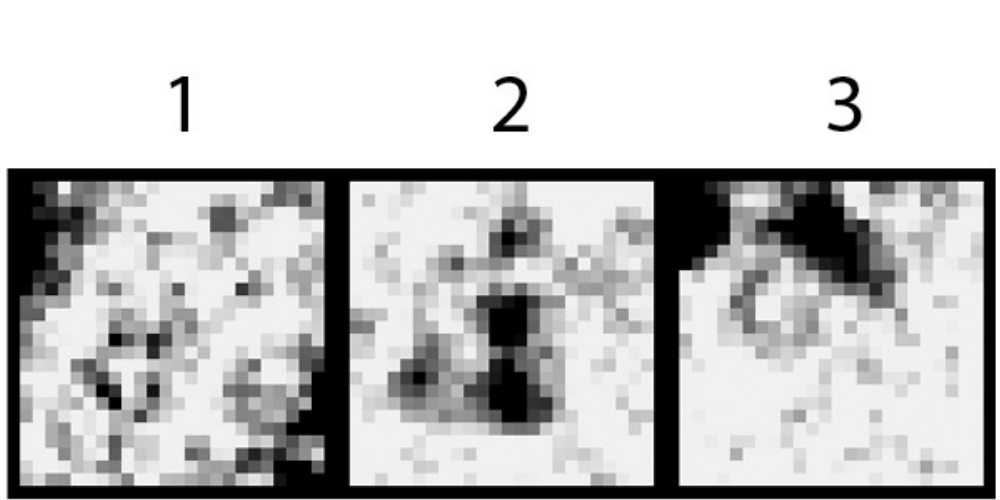} 
\caption{MIPS $24\mu$m postage stamps of the three candidate objects covering $30^{\prime\prime}\times 30^{\prime\prime}$.  Note only object 2 is detected with a flux of $66\pm12\mu$Jy. \label{mips-stamps}}
\end{center}
\end{figure}

\begin{deluxetable*}{ccccccccccccccc}
\tabletypesize{\scriptsize}
\tablecaption{Photometry \label{t:objects}}
\tablehead{
\colhead{ID} & \colhead{i$^+$}& \colhead{z$^+$}& \colhead{J}& \colhead{H}& \colhead{K}& \colhead{m3.6}& \colhead{m4.5}& \colhead{m5.8}& \colhead{m8.0}}
\startdata
1	& $>27.6$ & $>25.5$ & $23.21\pm0.05$ & $23.09\pm0.12$ & $22.14\pm0.02$& $21.89\pm0.02$& $21.65\pm0.03$& \nodata & \nodata \\
2	& $26.3\pm0.5$ & $25.22\pm0.5$ & $22.81\pm0.1$ & $23.00\pm0.11$ & $22.11\pm0.05$&$20.96\pm0.01$& $20.67\pm0.01$& $21.12\pm0.07$& $20.74\pm0.14$\\
3	& $>27.6$ & $>25.9$ & $23.11\pm0.1$ & $>23.5$ & $22.80\pm0.14 $&$22.34\pm0.03$& $22.16\pm0.05$& $21.9\pm0.1$& $22.2\pm0.7$\\
\enddata
\end{deluxetable*}

\begin{figure*}
\includegraphics[scale=1.08,trim= 0 0 0 0 0 ]{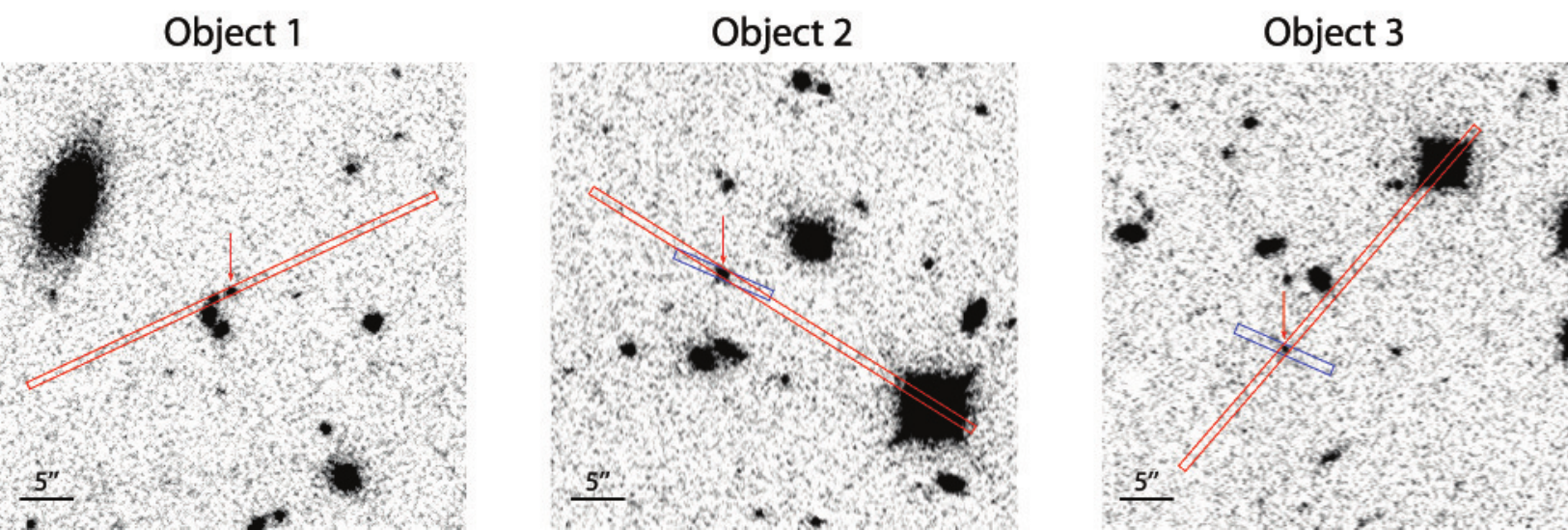}
\caption{\label{f:finder} The positions of the NIRSPEC (red) and DEIMOS (blue) slits are overlaid on the CFHT $K_s$ band images with North up and East to the left.  The targeted object is marked with a red arrow.}
\end{figure*}

	A $K_s$ band selected object catalog was generated with SExtractor \citep{1996A&AS..117..393B} using the CFHT-WIRCAM $K_s$ image for detection and PSF matched images for photometry as described in \cite{capak-data}.  The SExtractor detection settings are those used in \citet{McCrakenK}.  The $K_s$ band image was chosen over $J$ band for detection because while the image depths are similar, the image quality is significantly better in the $K_s$ band images.  Moreover, the expected object colors mean they should be equally bright or brighter in $K_s$ than $J$.  In addition to a detection in $K_s$ band, we also require a $5\sigma$ measurement in a 3\asec aperture for both the $J$ and $K_s$  bands for an object to enter our analysis.  This selection improves the robustness of our object identification, removing a significant number of artifacts which are only present in either the $J$ or $K_s$ band images.  The resulting catalog has a  $5\sigma$ limiting magnitude in a 3\asec aperture of $23.7$ on the AB system and is 80\% complete at this flux level for point sources  \citep{McCrakenK}.

	The $H$ band data was not available when the objects were selected, but is used in later analysis.  The photometry was measured in the same way as the for the $J$ and $K_s$ bands and has a $5\sigma$ limiting magnitude in a 3\asec aperture of $24.0$ on the AB system.
	
	As our primary selection criteria we require:  $J<23.7$, $z^+ - J \geq 1.5$, $J - K_s > 0$, $K_s - 4.5\mu$m$>0$, no bad pixels or cross talk in $z^+$ or $J$ bands, and less than a $2\sigma$ detection in $B_J$, $g^+$, and $V_J$ bands.   This selection removes many galactic stars which have blue $J - K_s$ and $K_s - 4.5\mu$m$>0$ colors and most low-redshift galaxies which are detected in the optical bands.  However, our sample still contained over 3500 objects, many of which appear to be dusty or passive $z\sim2$ galaxies or galactic L and T dwarf starts based on their spectral energy distributions (SEDs) (see Section 5).  Postage stamps in all 22 bands and a weighted average combination of the $0.3-0.8\mu$m bands were visually inspected to search for low level optical flux, resulting in a sample of 22 objects with no clear optical detections.  Of these 22 objects, 3 had flat SEDs from 1.2-4.5$\mu$m,  consistent with $z>7$ objects.  The remaining 19 candidates have very red SEDs implying they could be dusty $z\sim2$ galaxies.   The positions of the objects are given in Table \ref{t:objpos} and the $F814W$,  and $z^+$ limits along with measured fluxes for the 3 objects selected for follow-up are tabulated in Table \ref{t:objects}.   The 2$\sigma$ limiting fluxes in a 3\asec aperture for the other broad band images are $u^*>27.3$, $B_J>27.9$, $g^+>27.2$, $V_J>27.1$, $r^+>27.3$, and $i^+>26.9$.  The intermediate and narrow bands have limiting magnitudes $\sim1-2$ mag brighter than the broad bands \citep{2009ApJ...690.1236I}. 
		 		
	The color-color plot for the selected objects are compared with those from the literatures in Figure \ref{f:colors}.  Postage stamps of the ground based 0.3-2.2$\mu$m images along with Spitzer IRAC Ch1 (3.6$\mu$m) and Ch2 (4.5$\mu$m) are shown in Figure \ref{stamps}.  HST-ACS stamps are shown in Figure \ref{acs-stamps}.  All three objects have Full Width at Half Max (FWHM) measurements in the $K_s$ band which are larger than that measured for nearby stars.  However, only object 3 is inconsistent with a star based on the SExtractor stellarity index which uses a neural network to consider the effects of signal-to-noise and confusion on the size measurements.   Object 1 is near several foreground, optically detected sources which partially confuse the IRAC photometry for this source.  In the 3.6 and 4.5$\mu$m bands fluxes were recovered using 1.4\asec diameter apertures centered on the $K_s$ band position and subsequently correcting for the missing flux.  However, the fluxes at 5.8 and 8.0$\mu$m could not be reliably measured without significant de-convolution. Objects 2 and 3 are isolated.  Although object 2 is technically detected in the ground based $i^+$ image we included it in our sample because the detection is marginal and it is not detected at $>2\sigma$ in the ACS F814W data or the $z^+$ band data.  
	
	None of the sources is detected in the XMM 0.5-10 keV band X-ray data at a limit of $7.26 \times 10^{-16}$ ergs s$^{-1}$ cm$^{-2}$.  Sources 2 and 3 fall into a region with available Chandra 0.5-10 keV band x-ray imaging and are not detected at a $3\sigma$ upper limit of $7.4\times10^{-16}$ and $1.0\times10^{-15}$ ergs s$^{-1}$ cm$^{-2}$  respectively. 
	
	Source 2 is detected in the deep MIPS GO-3 imaging of the cosmos field with a flux of $66\pm12\mu$Jy at 24$\mu$m \citep{2009ApJ...703..222L, sanders-spitzer}.  Sources 1 and 3 are undetected in the GO-3 MIPS 24$\mu$m images.  Postage stamps of the Spitzer MIPS 24$\mu$m imaging are shown in Figure \ref{mips-stamps}.
	
	Sources 2 and 3 fall into a region with deep MAMBO \citep{2007ApJS..172..132B} imaging at 1.2mm, and neither is detected with fluxes $>2$mJy.  Finally, none of the sources are detected in the VLA 20 cm data with an RMS of $16\mu$Jy for object 1 and $10\mu$Jy for objects 2 and 3 \citep{eva-vla}  or at 90 cm with an RMS of 0.53, 0.42, and 0.47 mJy for objects 1, 2, and 3 respectively (Smolcic et al. in prep).
	
\section{Spectroscopic Observations Of Candidate Objects \label{s:spec}}

\begin{figure}
\includegraphics[scale=2,trim= 0 0 0 0 0 ]{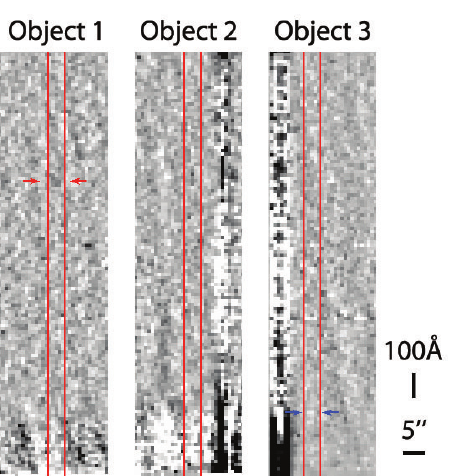}
\caption{\label{f:nirspec-spec1} The Keck NIRSPEC N1 spectra covering 0.9412-1.1097$\mu$m is shown with the object positions bracketed by vertical red lines which are 3.8\asec apart.   The greyscale is set so that darker is more positive. For object 1 note the weak line detection at 1.0563$\mu$m marked by a red arrow  which would place it at $z=7.69$ if this is Ly-$\alpha$, in good agreement with the photometric redshift.  There is no clear detection in the object 2 spectra.  For object 3 the position of the line seen in the DEIMOS spectra is indicated with a blue arrow but there is no detection in the NIRSPEC spectra, however the DEIMOS spectra is $\sim6\times$ as sensitive at the NIRSPEC spectra at this position.}
\end{figure}

\begin{figure}
\includegraphics[scale=2,trim= 0 0 0 0 0 ]{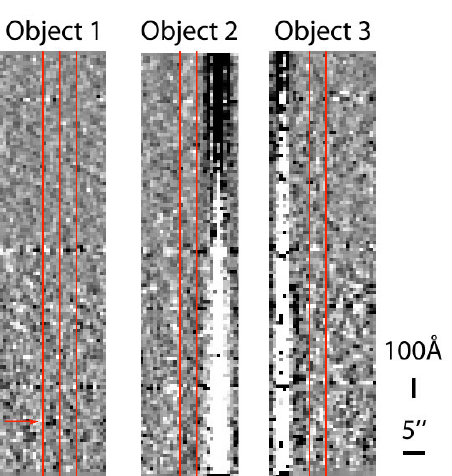}
\caption{\label{f:nirspec-spec5} The Keck NIRSPEC N5 spectra covering 1.5291-1.7953$\mu$m is shown with the object positions bracketed by red vertical lines which are 3.8\asec apart.  The greyscale is set so that darker is more positive. For object 1 the object position box is split in half with the foreground object on the left, and the target object on the right.  Note the line in the foreground object spectra at 1.560$\mu$m indicated by a red arrow.  This places the object at $z=1.37$ if this is H$-\alpha$.  There is no clear detection in the object 1 or  3 spectra.}
\end{figure}

	Near-IR spectra were obtained on February  22 and 23, 2008 under photometric conditions with 0.5\asec-0.7\asec seeing using NIRSPEC on the Keck-II telescope \citep{McLean:1998p702}.  The data were collected in low-resolution long slit mode using the 0.76\asec slit with the N1 and N5 order blocking filters.  The resulting spectral resolution was $R\sim1500$ with a pixel scale  $\sim1.6\AA$ and $\sim2.7\AA$ in  the N1 and N5 setting respectively and a spacial pixel scale of 0.18\asec.  The grating was tilted to cover 0.94-1.10$\mu$m and 1.52-1.80$\mu$m in N1 and N5 respectively.    These ranges were chosen to cover Ly$-\alpha$ (1216\AA) at $6.7<z<8$, [O II] (3727\AA) at $1.5<z<2$, and H$-\alpha$ (6563\AA) at $1.3<z<1.8$.   A nearby bright star was used to guide the acquisition.  For objects 2 and 3 the slit PA was set so we could observe the alignment star and the object simultaneously in at least one of the dither steps.    The data were taken in an "ABBA" dither pattern with a 15\asec throw along the slit.  The integration times for objects 1, 2, and 3 respectively are 2h, 3h, and 3.75h in N1 and 1h for all objects in N5.  The data were reduced using the NIRSPEC reduction package described in \citet{2006ApJ...640...69B}.  Once reduced, the slits were rectified and co-added using a bright object positioned on the slit to register the images.
	
	Latent images from prior observations can be a problem in NIRSPEC data.  To ensure these were not affecting the final results we inspected each frame for latent images.  We also split our data by night, dither pattern, and dither position, inspecting each of these stacks for latent images which appeared in one frame but not others.  Latent images from a bright standard used to determine the slit function were only visible in the first dither N1 dither pattern of the spectra taken on object 3 on Feb 22 and the first N1 exposure taken on object 3 on Feb 23.  The latents were not at the expected object position.

\begin{figure*}
\includegraphics[scale=0.8,trim= 0 0 0 0 0 ]{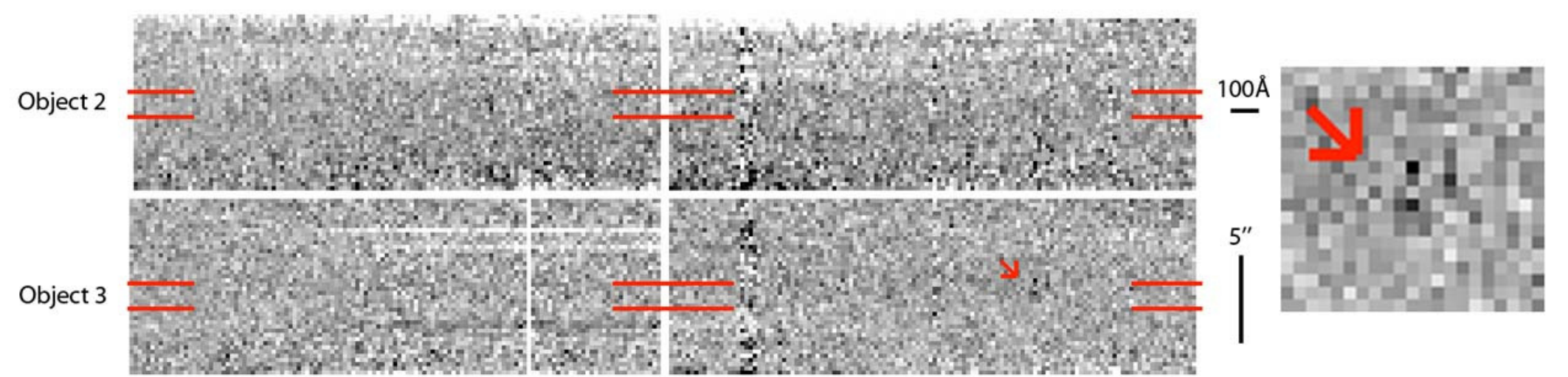}
\caption{\label{f:deimos-spec} The Keck DEIMOS spectra covering 0.64-1.02$\mu$m are shown with the object positions bracketed by red horizontal lines at the end of each spectra.  The greyscale is set so that darker is more positive. Note the possible line detection in object 3 at 0.9668$\mu$m indicated by the red arrow and enlarged in a panel at the right of the figure.  This suggests a redshift of 6.95 if the line is Ly-$\alpha$. }
\end{figure*}

\begin{deluxetable*}{cccccc}
\tabletypesize{\scriptsize}
\tablecaption{Line Sensitivities and un-obscured SFR sensitivities\label{t:sensitivity} }
\tablehead{
\colhead{ID} & \colhead{0.6-1$\mu$m} & \colhead{0.95-1.1$\mu$m}      & \colhead{1.5-1.8$\mu$m}&  \colhead{SFR-[O II] (M$_\odot$ yr$^{-1}$)}& \colhead{SFR H-$\alpha$ (M$_\odot$ yr$^{-1}$)} \\
\colhead{} & \colhead{erg cm$^{-2}$ s$^{-1}$} & \colhead{erg cm$^{-2}$ s$^{-1}$}      & \colhead{erg cm$^{-2}$ s$^{-1}$}& \colhead{$1.0<z<2.0$, $3.1<z<3.8$}& \colhead{$1.3<z<1.7$}
}
\startdata
1	&	\nodata &  $11-122 \times 10^{-18}$ & $1-10 \times 10^{-17}$ & 1-49, 12-190 & 1-16 \\
2	&	$1-10 \times 10^{-18}$  &  $9-100 \times 10^{-18}$ & $1-10 \times 10^{-17}$ & 1-40, 12-190 & 1-16 \\
3	&	$1-10 \times 10^{-18}$ &  $6-70 \times 10^{-18}$ & $1-10 \times 10^{-17}$ & 1-28, 12-190 & 1-16\\
\enddata
\end{deluxetable*}

	Optical spectra were obtained for objects 2 and 3 with DEIMOS on the Keck-II telescope \citep{Faber:2003p717} on November 23, 2008 under partially cloudy conditions and November 25, 2008 under clear conditions with $\sim1$\asec seeing. The data were collected with 1.0\asec wide slits, the the 830l/mm grating tilted to 7900\AA\, and the OG550 blocker.  The resulting spectral resolution was $R\sim2370$ with a pixel scale of $\sim0.47\AA$ a spacial pixel scale of 0.1185\asec.  The data were reduced with a modified version of the DEEP2 DEIMOS pipeline \citep{2001astro.ph..9164M}.  In addition to the standard processing this modified pipeline constructs and subtracts a median background and accounts for dithering, positional shifts, and photometric scaling when combining the spectra.  The objects were dithered $\pm3$\asec along the slit to improve the sky subtraction and mitigate ghosting (internal reflections) from the 830l/mm grating.  Bright objects in the mask were used to determine the amount of atmospheric extinction on November 23, and integrations with less than 90\% throughput were removed.  Each exposure was 30min for a total of 4.5h of un-extincted integration time.

	The positions of the NIRSPEC and DEIMOS slits are shown in Figure \ref{f:finder} and the 2D spectra are shown in Figures \ref{f:nirspec-spec1}, \ref{f:nirspec-spec5}, and \ref{f:deimos-spec}.  We tabulate the range in line sensitivity assuming a 10\AA\ extraction box in the spectral domain, comparable to the expected observed frame line width, and the limits this places on the amount of star formation in Table \ref{t:sensitivity}.

	We conducted a close inspection of the spectra, including smoothing and binning by up to 32 pixels ($\sim 115$\AA) in the dispersion direction and 10 pixels (1\asec) in the slit direction.  Since the visible spectral features are weak we used two methods to estimate their significance.  First, we created a variance map including read noise, Poisson noise, estimated errors in background subtraction, variations in depth from bad pixels and dithering, and noise correlation due to re-sampling.  This map was used to estimate the significance of any feature.  As a second estimate of significance we used SExtractor to search for low signal-to-noise peaks in the 2d spectra with the object and alignment star positions masked.  Using this catalog of noise peaks, we then calculate the probability that a peak of a given flux would fall within $\pm1\asec$ of the object position.   
	
	At the expected position for object 1 a weak line is visible at 1.0563$\mu$m in the N1 spectra.  The significance of this feature is $2.4\sigma$ based on the noise map and $2.6\sigma$ based on the noise peak method.  If this line is Ly-$\alpha$ it places the object at $z=7.69$, however the line could also be [O II] at $z=1.83$.  Nothing is seen in the N5 spectra of object 1 at the expected position. However, a line is seen at 1.560$\mu$m at the position of a nearby optically detected foreground object seen in Figure \ref{stamps} and noted in Section 2.  The noise map yields a significance of $3.8\sigma$ for this feature and no other peaks at this flux level are found in the spectra.  This would place the neighboring object at  $z=1.37$ if the line is H$-\alpha$, which is the solution favored by the photometric redshift from \citet{2009ApJ...690.1236I}, but  [O II]  at $z=3.19$ can not be ruled out.   [O III] (5007\AA) or H-$\beta$ (4861\AA) lines at $z\sim 2.2$ are unlikely for both objects because we would expect to see both lines in the spectra.  However, we note there is significant scatter in the observed [O III] and H-$\beta$ line ratios so this redshift can not be entirely excluded at our signal-to-noise.

	Nothing is detectable in the DEIMOS, N1 or N5 spectra of object 2.  
		 	
	The DEIMOS spectra of object 3 shows a feature at 0.9672$\mu$m with a $3.4\sigma$ significance based on the noise map and no other features with equal significance found in the spectra.  This suggests a redshift of 6.95 if the line is Ly-$\alpha$, assuming the blue edge is at the line center.  However this line could also be [O II] at z=1.60 or less likely Ly-$\beta$ at z=8.43.  Nothing is detected in either the N1 or N5 spectra for object 3.  The sensitivity of the N1 spectra is $\sim6\times$ lower than the DEIMOS spectra at $\sim 0.9672\mu$m so the non-detection does not rule out the line seen in the DEIMOS spectra.  The sensitivity in the N5 spectra is also insufficient to detect H$-\alpha$ at $z=1.60$ for the typical H-$\alpha$  to [O II] star forming line ratio of 1.77 \citep{1998ARA&A..36..189K} unless the source were obscured by A$_v>7$, which is significantly  larger than the value implied by the broad band colors.
		
\section{SED Fitting Analysis \label{s:SED}}

\begin{figure*}
\begin{center}
\begin{tabular}{ccc}
\includegraphics[scale=0.3]{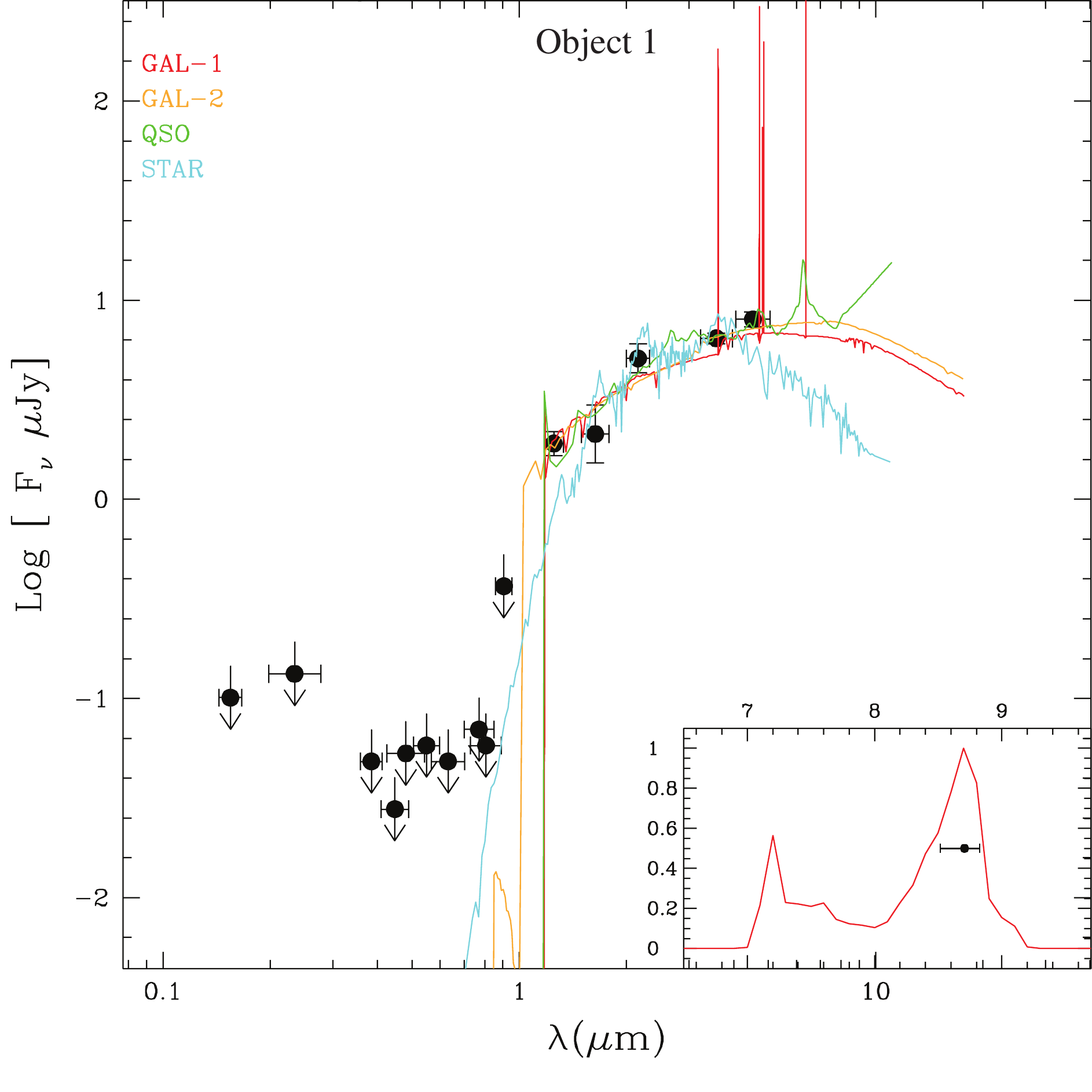}& \includegraphics[scale=0.3]{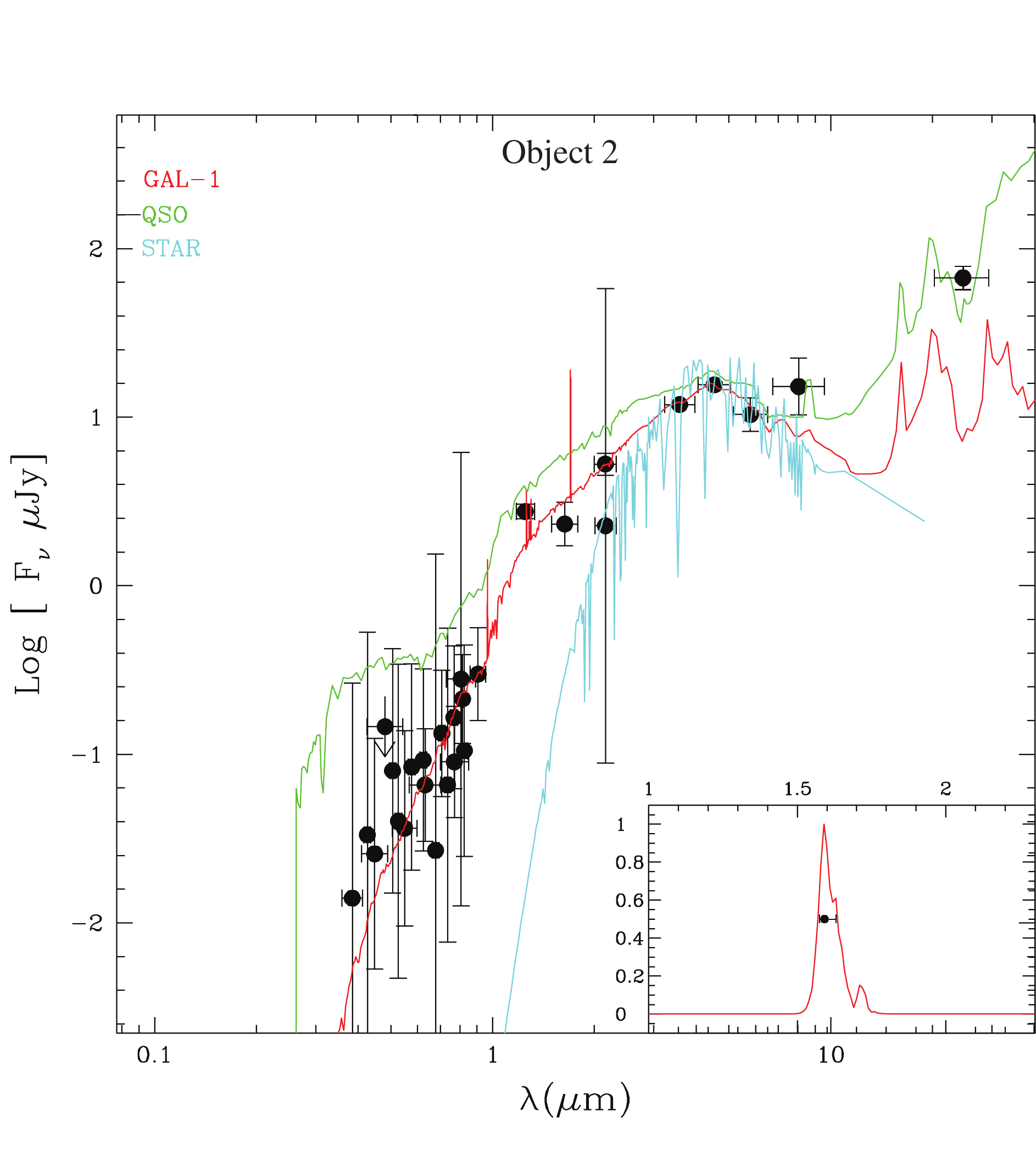}& \includegraphics[scale=0.3]{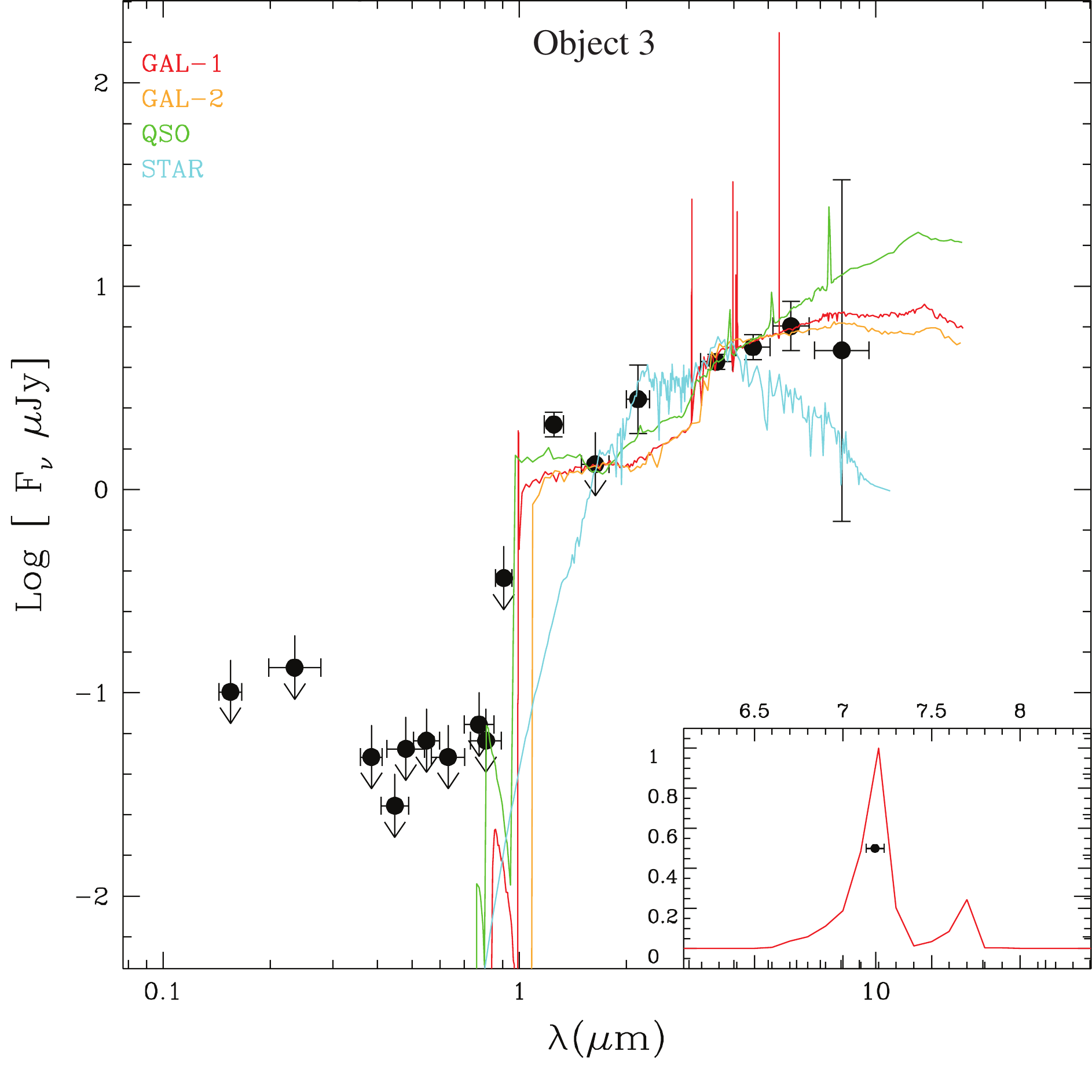} \\
\end{tabular}
\caption{The best fit SEDs and redshift probability distributions are shown for all three objects.   The best fit galaxy SED is shown in red and a secondary galaxy fit, if one exists, shown in orange.  The best fit Quasar template is shown in green and the best fit stellar spectra is shown in cyan.  The joint probability distribution for all SED types is shown in the bottom right corner of each box for regions where the probability is greater than 0.01\%.  Note the lack of any low redshift solution for object 1 and 3  and the peak at $z \sim 1.5$ for object 2.  The templates are empirically extended at rest frame wavelengths greater than $2\mu$m.  \label{f:SED}}
\end{center}
\end{figure*}

	To quantify the probability these sources are at $z>7$ and constrain their physical properties, a SED fitting analysis was conducted using the {\it Le Phare} code \footnote{http://www.oamp.fr/people/arnouts/LE\_PHARE.html}.  Both empirical \citep{Polletta:2007p289, Budavari:2001p404} and theoretical \citep[][2007]{2003MNRAS.344.1000B} templates were fit with up to $A_V=3$ mag of obscuration using a \citet{1996ApJ...458..132C} or \citet{Prevot:1984p381} extinction law and an additional 2175\AA\ bump added as a free parameter to the extinction law \citep{2009ApJ...690.1236I}.  Stars from the \citet{Bixler:1991p444} and \citet{Chabrier:2000p429} libraries covering O to T stars were also fit to the photometry.  The optical detection limits were conservatively set at $3\sigma$ rather than $1\sigma$ values since our knowledge of $z>1$ SEDs are limited.  The results of the SED fitting analysis are shown in Table \ref{t:SED} and Figure \ref{f:SED}.

	The $J$ band flux is critical in the SED fitting analysis because it constrains the break strength between the $z^+$ and $J$ bands, which in turn constrains the redshift \citep{2008MNRAS.386..370S,2008ApJ...686..230B} .  However, this is mitigated somewhat by the high-quality Spitzer IRAC and CFHT $H$, $K_s$ band data which also place strong constraints on the SED shape. To test for the effects of photometric offsets we introduced a shift of $\pm0.5$ magnitude in $J$ band  and removed it and the IRAC 5.8 and 8.0$\mu$m bands from the fits.

\begin{deluxetable}{cccccc}
\tabletypesize{\scriptsize}
\tablecaption{SED Fitting results\label{t:SED}}
\tablehead{
\colhead{ID} & \colhead{SED Type} & \colhead{Redshift} & \colhead{e(B-V)}  & \colhead{Mass} & \colhead{Age} }
\startdata
1	&	Emperical &	$7.6^{+0.3}_{-0.5}$       & 0.3 & \_ & \_  \\
2	&	Emperical &	$1.59^{+0.07}_{-0.02}$       & 0.5 & \_ & \_ \\
3	&	Emperical &	$6.86^{+0.1}_{-0.1}$       & 0.3 & \_ & \_ \\
1	&	BC03 &	$7.7^{+1.2}_{-0.4}$ & 0.4 & $1\times10^{11}$ & 4-30 Myr \\
2	&	BC03 &	$1.59^{+0.07}_{-0.02}$ & 0.5 & $4\times10^{11}$ & 4 Myr  \\
3	&	BC03 &	$7.2^{+0.1}_{-0.2}$ & 0.1 & $2\times10^{11}$ & 450-600 Myr  \\
\enddata
\end{deluxetable}

		The best fit for object 1 is an obscured (A$_v=1.2$) star forming galaxy or un-obscured AGN at $7.1<z<8.9$, in good agreement with the potential line at $z=7.7$.  The star formation rate based on the SED fit would between 190-3000 M$_\sun yr^{-1}$, depending on the obscuration correction, so this source may be detectable in the millimeter if it is a star forming galaxy. If instead the source is a Type 1 AGN, the x-ray flux limits the black hole mass to $<6\times10^7$M$_\sun$ at $z\sim7$ based on an analysis of Type 1 Quasars in the COSMOS field (Lusso et al., submitted).
	
	Removing or decrementing the $J$ band magnitude by 0.5 mag for object 1 produces photo-z solutions between $0.6<z<1.2$ with an extinguished star forming galaxies.  However, all of these fits imply $>3\sigma$ detections in the $i^+$ and $F814W$ bands and $>2\sigma$ detections in the $z^+$ band.  Furthermore, they suggest star formation rates of 30-1000 M$_\sun yr^{-1}$, which should lead to a MIPS 24$\mu$m detection \citep{2009ApJ...692..556R}.  As a result these low-z solutions are strongly disfavored by the data and it is likely that this object is at $z>7$.

	For object 2 the low S/N $i^+$ and $z^+$ band detections and 24$\mu$m detection firmly place it at low redshift.  The best fit SED  is a very young, obscured (A$_V$=1.55) source at $z=1.59$ with 15 M$_\sun yr^{-1}$ of star formation, in broad agreement with the 24$\mu$m flux.

	The best fits for object 3 place it at $6.7<z<7.3$ with a 200Myr old relatively un-obscured (A$_v=0.3$) stellar population or an AGN type spectra.  The star formation rate implied by the SED fit is $\sim200$ M$_\sun yr^{-1}$, comfortably within the mm detection limits.  However the weak $H$ band flux relative to the $J$ and $K$ is a quandary.  This could be explained by an unusually strong 2175\AA\ dust absorption feature, but, evidence suggests that dust at these high redshifts does not have a strong 2175\AA\ feature \citep{2007ApJ...661L...9S,2004Natur.431..533M}.   Alternatively, the source could be an AGN at $z>8$ with Ly-$\alpha$ and CIV falling in the $J$ band and $H$ band covering a region of relatively line free continuum.  Following this interpretation the line seen in the DEIMOS spectra would be Ly-$\beta$ at $z=8.43$, which is also seen in deep spectra of the $z>6$ SDSS quasars \citep{2006AJ....131.1203F}, but would imply a very strong Ly$-\alpha$ line flux to overcome the strong absorption from the intergalactic medium at these redshifts.  If this is an AGN, the Chandra x-ray flux limits the black hole mass to $<6\times10^7$M$_\sun$ at $z\sim8.4$ (Lusso et al., submitted).  
		
	 Removing the $J$ band data point resulted in a fit at $2<z<4$ by an obscured (A$_v=0.3$) $\sim200$ Myr old stellar population with $\sim200$ M$_\sun yr^{-1}$ of star formation.  However, the fits imply $>3\sigma$ detections in the $i^+$ and $F814W$ bands and a $>5\sigma$ detection in the $H$ band which is not seen.  	

\section{Contamination from $z<<7$ \label{s:contam}}

	The reliability of photometric redshifts is tightly linked to how well the input SEDs represent the galaxy population.  Photometric redshifts and photometric selections have thus occasionally erroneously identified exotic low-redshift objects as high-redshift systems \citep{2008ApJ...680L..97H, 2005ApJ...635..832M, 2004A&A...416L..35P,2000ApJ...531..624D}.  For $z>7$ galaxy searches this is especially worrisome in view of the large numbers of optically faint galaxies in the "redshift desert" ($1<z<2.5$) \citep{2004ApJ...604..534S,2004ApJ...617..746D} and late type stars which have similar colors and poorly constrained SEDs.  These contaminating objects are difficult to study for the same reasons as $z>7$ objects; their strong spectral features are shifted to NIR wavelengths where the night sky is brighter, the density of night sky lines higher, and instruments have a smaller fields of view and lower sensitivity.  
	
\begin{figure}
\includegraphics[scale=0.36,trim= 0 0 0 0 0 ]{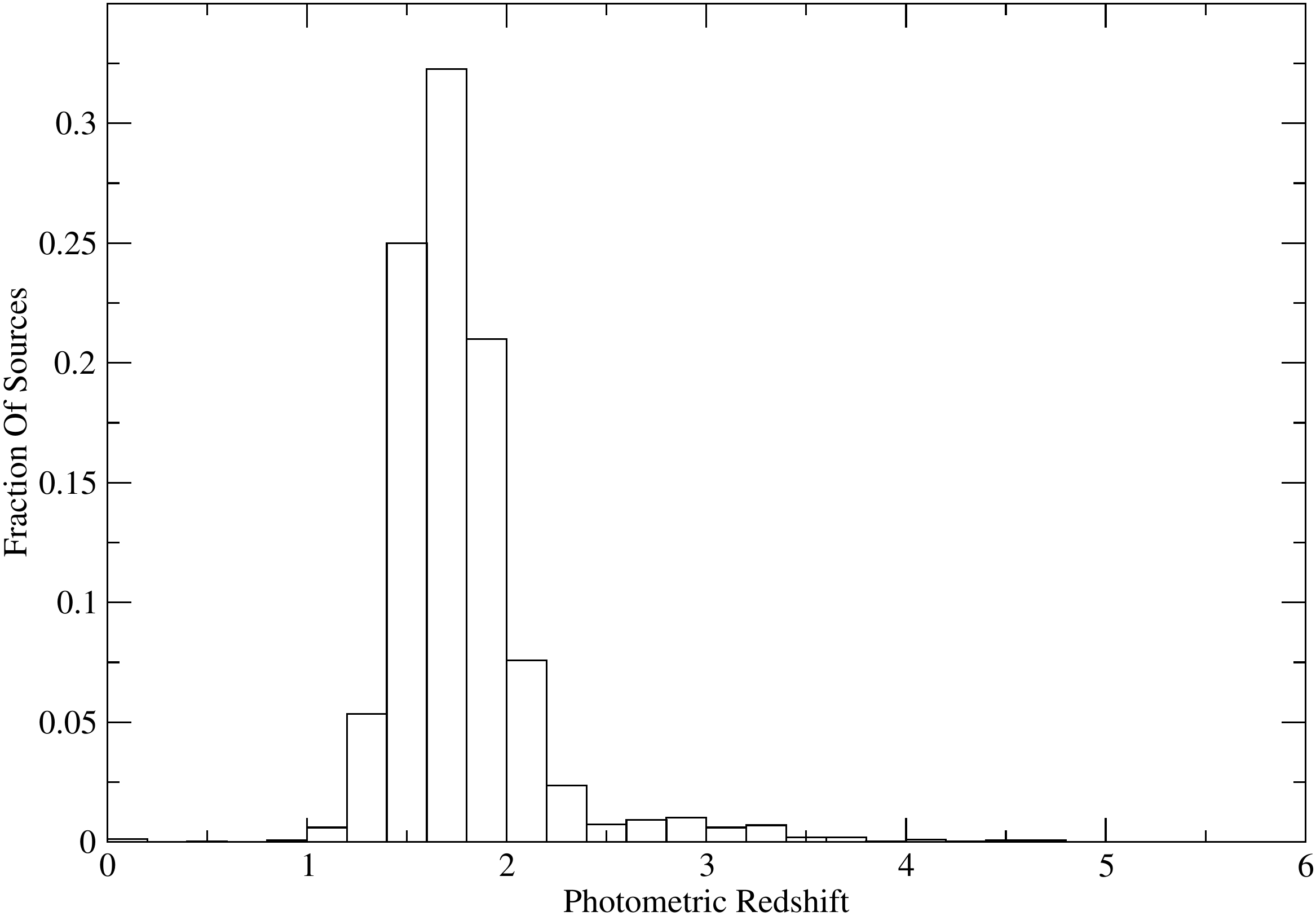}
\caption{\label{f:contam-zdist}The photometric redshift distribution from \citet{2009ApJ...690.1236I} for $z<<7$ objects selected by a $z^+-J > 1.5$, $J-K >  0$, $K_s - 4.5\mu$m$>0$ color cut is shown. Note the peak at $z\sim1.8$ where the 4000\AA\ break and dust obscuration is confused with Lyman-$\alpha$ forest absorption at $z\sim7$.}
\end{figure}

\begin{figure}
\includegraphics[scale=0.34]{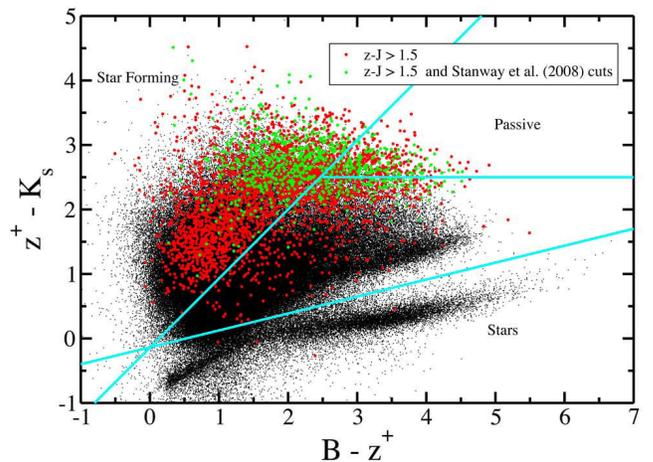}
\caption{\label{f:bzk}The $(B-z^+)$, $(z-K_s)$ color-color diagram for an $i$ band selected sample is shown.  Objects with $z^+ - J > 1.5$ colors are marked in red, and those detected in all 4 IRAC bands meeting the NIR and IRAC color cuts suggested by \citet{2008MNRAS.386..370S} are plotted in green.  The position of the sources meeting the z-drop criteria indicate they are obscured star forming galaxies and passive galaxies at $z\sim2$.}
\end{figure}

	To understand the limitations of our selection and the SED models we selected a sample of optically bright ($i^+<26.5$) objects from the COSMOS $i$-band catalog \citep{capak-data} which meet our $z^+$ band dropout selection criteria ($z^+ - J > 1.5$, $J-K >  0$, $K_s - 4.5\mu$m$>0$).  The intermediate band data provides a low resolution spectra (R$\sim20$) for these bright objects, allowing us to classify them and obtain reasonable photometric redshifts as shown in Figure \ref{f:contam-zdist}.  The majority of these objects have photometric redshifts in the $1.4<z<2.5$ range (Figure \ref{f:contam-zdist}), so a BzK diagram (Figure \ref{f:bzk}) provides information on the properties of these galaxies \citep{2004ApJ...617..746D}.  The BzK diagram indicates a mix of obscured star forming and passive galaxies at $z\sim2$.  

\begin{figure}
\includegraphics[scale=0.35 ]{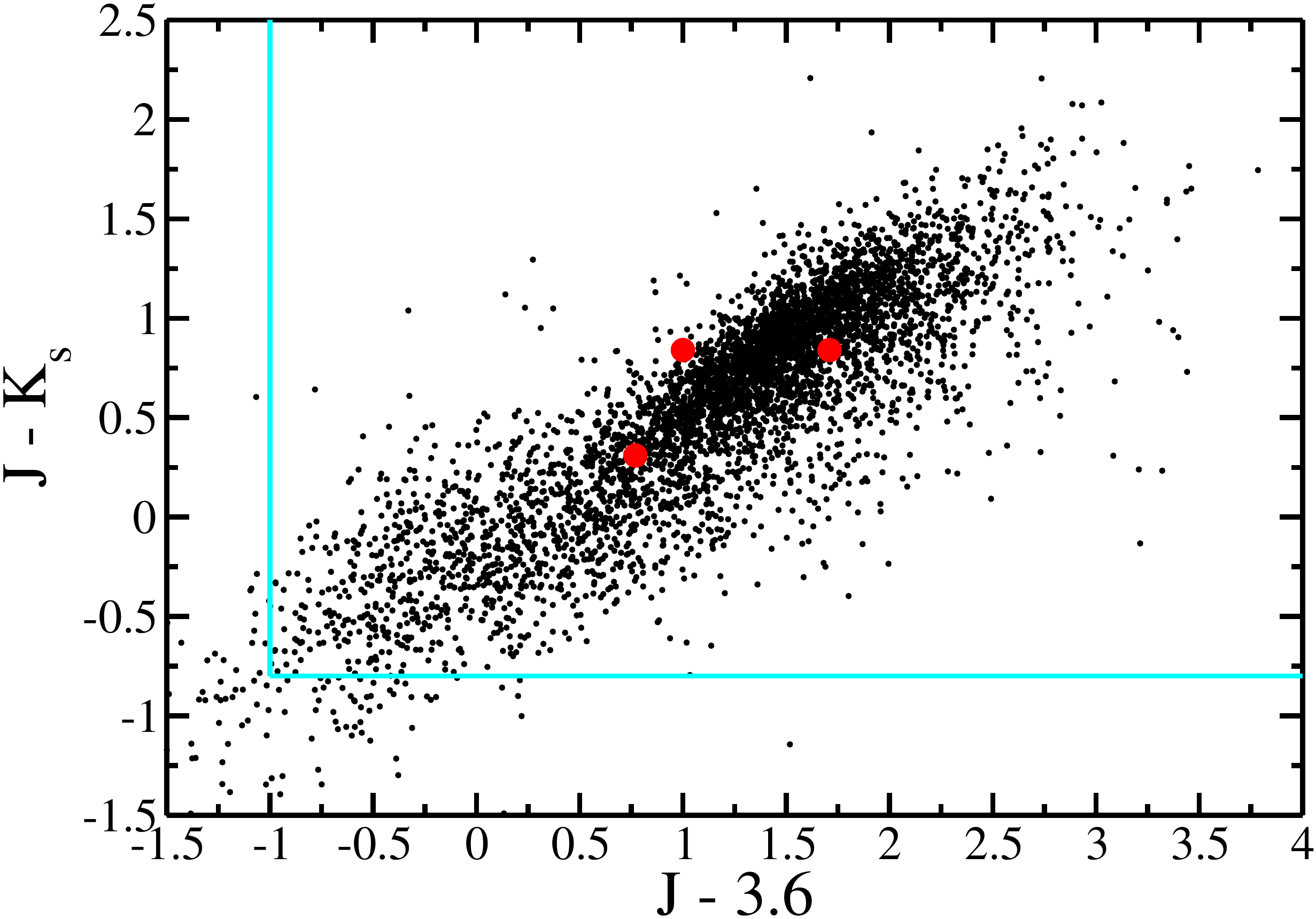}\\
\vskip 0.5in
\includegraphics[scale=0.35 ]{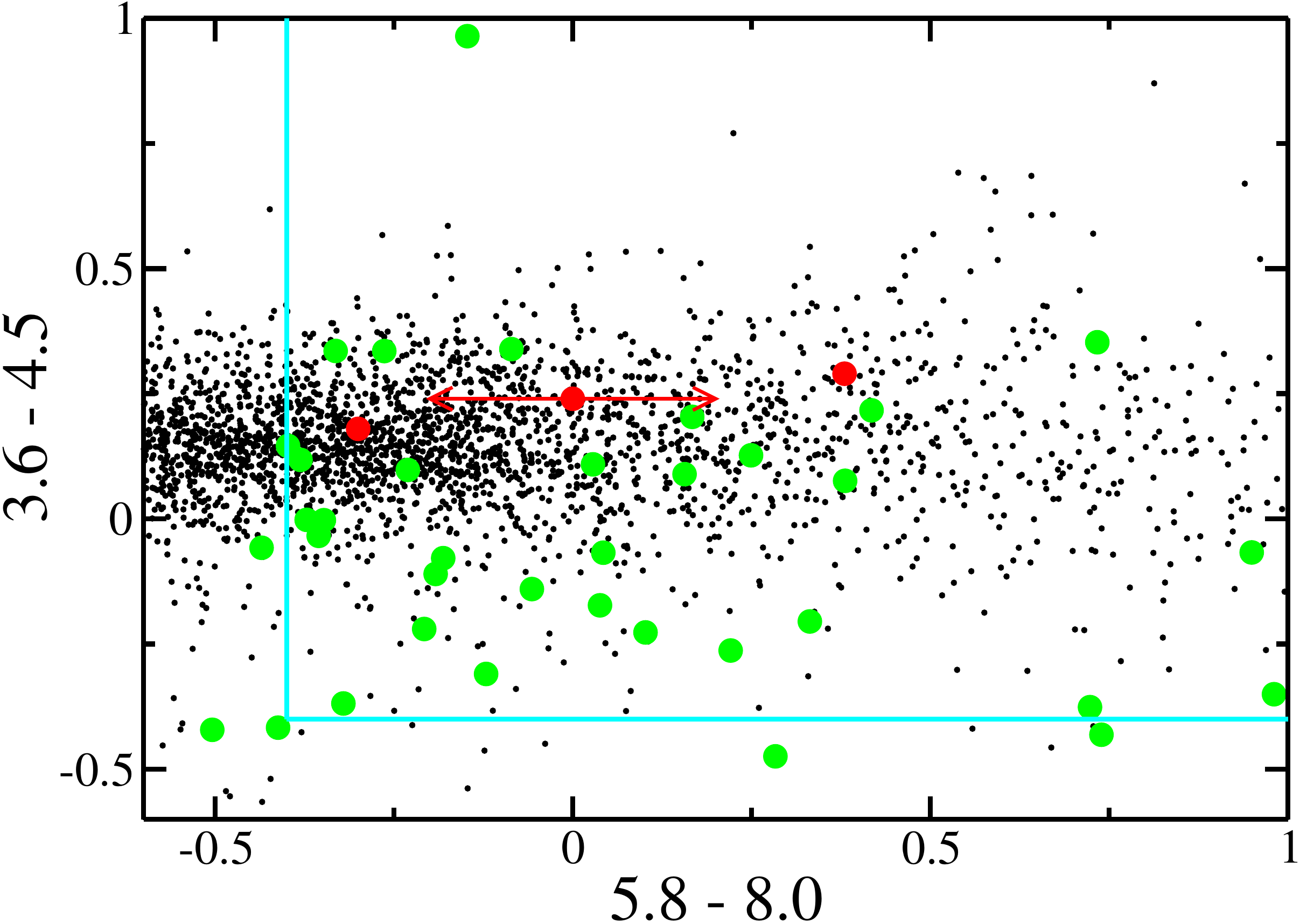}
\caption{\label{f:stanway}The NIR and IRAC colors of galaxies detected in i band with $z^+ - J > 1.5$ colors are plotted in black along with our candidate objects in red.  The color selections suggested by \citet{2008MNRAS.386..370S} are shown as cyan lines.  Objects removed by the criteria shown in the top panel are marked in green in the bottom panel.  Note a significant number of objects meet both color criteria but are not removed by the suggested color cuts.  These are likely obscured star forming and passive galaxies at $1<z<3$ (See Figures \ref{f:contam-zdist} and \ref{f:bzk}).}
\end{figure}

	Given this contamination we investigated the $J$, $K_s$ and IRAC color cuts proposed by \citet{2008MNRAS.386..370S}  (Figure \ref{f:stanway}) which were formulated to remove passive objects at $z<2.5$ and late type stars.  These cuts succeed in removing a large number of objects, but fail to remove a significant fraction of the contaminating population with the steepest spectral slopes.  These objects are likely to have a steeper dust extinction curve than the locally observed value \citep{2008ApJ...689...59S}  or be high redshift analogs of the dusty red galaxies seen in the A901/2 super cluster \citep{2009MNRAS.393.1302W}. 
	
	Another possible explanation are objects with extreme line equivalent widths such those presented in \citet{2007ApJ...668..853K,2009ApJ...698.2014H}.  However, at $z\sim1.8$ these objects would mimick the blue $J-H$ colors thought to be unique to young stellar populations at $z>7$ \citep{2010ApJ...709L..16O,2009arXiv0909.2255B}.  Furthermore, unlike other sources of contamination, these objects are likely to be more common at fainter magnitudes because they are thought to originate in low-mass, low-metalicity galaxies.  However, the same strong emission lines would also mean these sources could be easily identified with deep spectroscopy, if the sources are bright enough.

\begin{figure}
\includegraphics[scale=0.36]{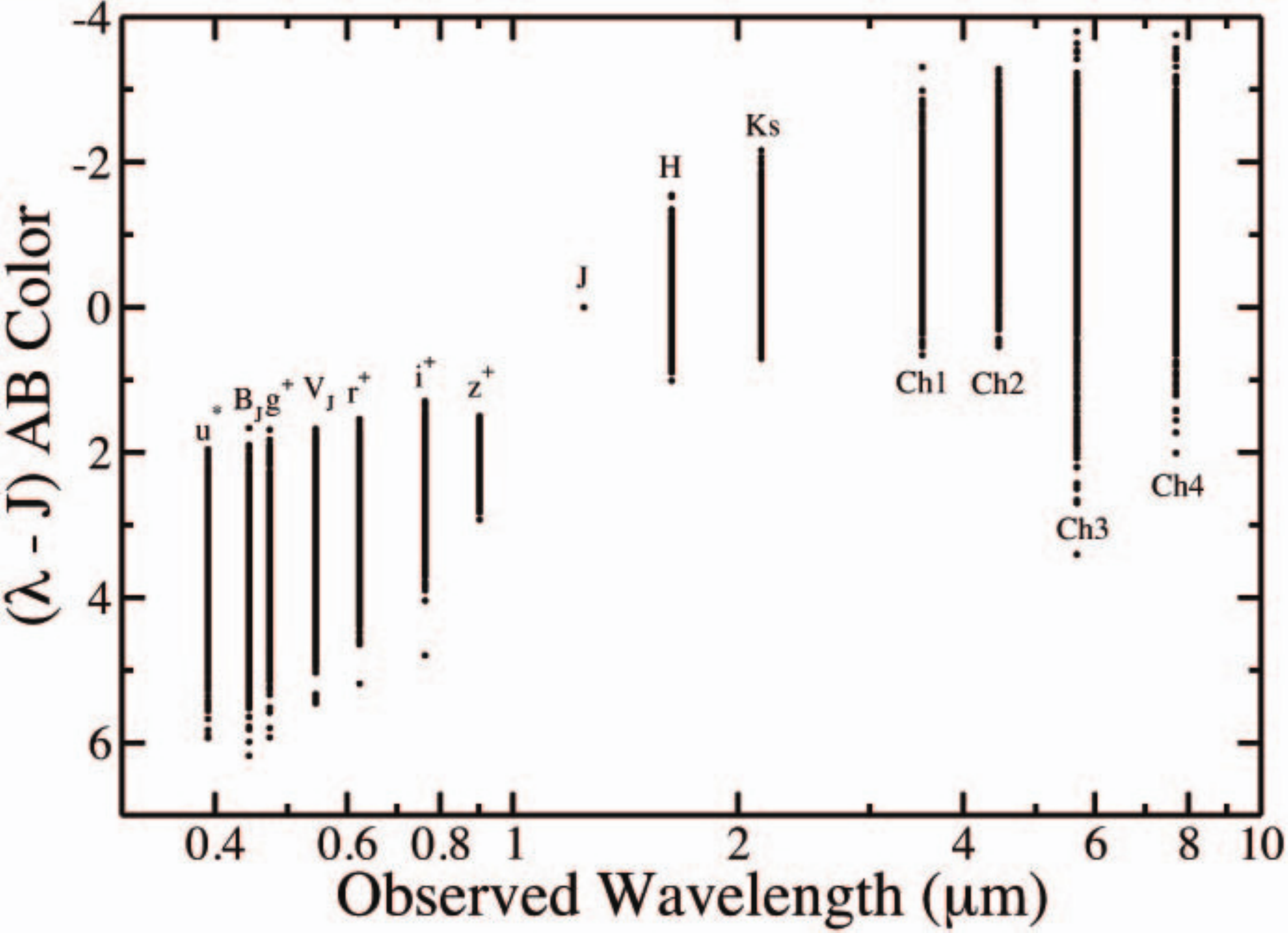}
\caption{\label{f:contam-phot} The (J-$\lambda$) colors of $z<<7$ objects selected by a $z^+-J > 1.5$, $J-K_s >  0$, $K_s - 4.5\mu$m$>0$ color cuts are shown.  Note that the $i$ band data must be 4 magnitudes deeper to remove these objects from the sample.}
\end{figure}

	Based on this analysis a range of NIR and IRAC color cuts can reduce the number of contaminating objects, but not completely remove them.  Furthermore, many of the $z<<7$ objects exhibit blue the $J-H$ colors thought to be indicative of young stellar populations at high redshift \citep{2010ApJ...709L..16O,2010ApJ...708L..69B,2009arXiv0909.2255B}. Therefore, blue NIR colors and a strong break are insufficient to confirm sources.    We conclude the strongest selection criteria presented here, and in the literature, are the non-detections in the $i^+$ and $z^+$ bands which rule out low-z objects based upon the strength and steepness of the detected break.   To explore the necessary depth in $i$ and $z$ bands required to rule out low-redshift objects, in Figure \ref{f:contam-phot}  we take the photometry for sources meeting our z-dropout criteria and scale them to a J band magnitude of zero .  This figure shows a significant number of low-z objects have $z^+-J > 2$ and $i^+-J > 3.5$ colors.  So a survey must be at least 3 magnitudes deeper in $z^+$ and 4 magnitudes deeper in $i$ than $J$ to adequately select $z>7$ galaxies. 

	Based on the limits required in $i$ and $z$ bands to rule out low redshift objects, most surveys have reached their maximum depth for $z>7$ searches unless deeper optical data is obtained.   Assuming $2\sigma$ limits are sufficient to rule out low-z objects, the COSMOS i band data are sufficiently deep to probe to $J=23.7$, the depth of our present survey.   The Hubble Ultra Deep Field (UDF) \citep{Beckwith:2006p497}  obtains $2\sigma$ depths of F775W=31.1, and F850LP=30.3, which is sufficient to remove contaminating objects brighter than F160W of 27.1,  well matched to the present NICMOS data. Finally, the Great Observatories Origins Deep Survey (GOODS) \citep{Giavalisco:2004p569} F775W and F850LP data reach $2\sigma$ depths of 28.9 and 28.6 respectively, sufficient for the depth of the present ground based NIR imaging with a limit of $J=24.9$ \citep{2007ApJ...670..928B}.  
		
	Relying on deep $i$ and $z$ band imaging is problematic as we probe deeper due to foreground objects which will occupy an increasing fraction of the sky.  This effectively reducing the sample completeness by removing sources that are superimposed on faint foreground sources by chance.  In our ground based COSMOS images we lose $>38$\% of the un-masked sky area to objects detected in the ground based optical bands.  The situation improves in space due to increased resolution and decreasing size of faint galaxies.  Assuming the size distribution as a function of magnitude given in \citet{2009A&A...504..359J}
, using the faintest magnitude bin to extrapolate to F775W=31.1 we find that $>10$\% of the UDF area will be lost to objects brighter than F775W=31.1. 

	

	Therefore, the $z-J$, $J-K$, and $F850LP-F110W$, $F110W-F160W$  based $z>7$ search techniques used from the ground and space may have reached there practical limits.  Deeper $z-J$ color searches with Ultra-Vista, the United Kingdom Infrared Deep Sky Survey (UKIDDS), or Wide Field Camera 3 (WFC3) will likely be dominated by contamination unless significantly deeper $i$ and $z$ band data are obtained, and even then they are likely to be incomplete due to confusion with foreground galaxies. However, narrower adjacent filter sets such as that used in the WFC3-UDF \citep{2010ApJ...709L..16O,2009arXiv0909.2255B} should decrease the amount of contamination by better differentiating the smooth roll-over in the SED of obscured and old stellar populations at $z\sim1.8$ and the sharp breaks caused by the Ly$-\alpha$ forest at $z>7$.  However, objects with strong line emission will still be problematic.

\section{Discussion}

\subsection{Are these objects at $z>7$?}	

	Based on the present photometric and spectroscopic data it is unclear whether or not the objects presented are genuinely at $z>7$.   All three objects meet  the color selections suggested by previous investigations \citep{2008MNRAS.386..370S, 2009ApJ...690.1350O, 2006ApJ...649L..67L, 2008ApJ...686..230B}.  However, the 24$\mu$m and marginal $i^+$ detection of object 2 strongly suggests it is an obscured source at $z\sim1.6$.  The SEDs of objects 1 and 3 are inconsistent with known $z<7$ galaxies, AGN or stars in our galaxy, and the features seen in the NIRSPEC and DEIMOS spectra strongly suggest objects 1 and 3 are at $z=7.7$ and $z=6.95$ respectively.
	
	  The broad band colors and SEDs of these objects are indistinguishable from $z>7$ candidates found in the UDF, GOODS and other deep surveys \citep{2009ApJ...690.1350O, 2008ApJ...686..230B}.  With a selection function similar to those of \citet{2009ApJ...690.1350O} and \citet{2008ApJ...686..230B} we estimate a space density of $2.51\pm1.78\times10^{-7}$ mag $^{-1}$ Mpc$^{-3}$ for objects with M$_{UV}=-23.85$ (see Figure \ref{f:lf-evolution}).  This space density is consistent with the faint end counts of $z\sim7$ galaxies from the WFC3-UDF data \citep{2010ApJ...709L..16O,2009arXiv0909.2255B} if the galaxy luminosity function follows the shape of the halo mass function shown as a black line in Figure \ref{f:lf-evolution}.  This is significantly higher than the space density implied by fits to smaller area data, but not above the theoretical limit for the density of these sources. 
	  
	 If these objects are at $z<<7$ they point to a population of stars and/or $1.4<z<2$ galaxies with unexpected colors that are contaminating our sample and the objects discovered in fainter samples may be contaminants.  This population of unusual objects must be understood before a reliable $z>7$ selection can be made.
	  	  
	We also explored the possibility that object 1 is strongly lensed rather than intrinsically bright.  To do this we constructed a 3D lens model with "Lens Tool"  \citep{2007NJPh....9..447J, 1993A&A...273..367K} using the stellar mass and photometric redshfits of the three nearby foreground objects .  We assumed a dark-matter to stellar-matter ratio of 20-1 and a Navarro-Frenk-White (NFW) \citep{1997ApJ...490..493N} profile scaled to the observed light profiles of the galaxies.  This produced a magnification of $\sim 1.2$, and up to $\sim 2$ if the galaxies were several times more massive than estimated.  As a result, we believe lensing magnification is an insignificant  contribution.
	 
	The lack of conclusive spectra is troubling because these objects are at the practical spectroscopic limit for 8-10m ground based telescopes.  Integration times of $\sim 25$h ($\sim3$ nights) per object would be required to reach a S/N of 5 for the features seen in these spectra.  Multi-band NIR ground based spectrographs will allow better probes by simultaneously recording spectra from a broad wavelength range, however gaps will sill remain in the spectral coverage due to atmospheric absorption.  Multi-object spectrographs will not provide a significant advance because the sky surface density of these sources is low, even at fainter magntitudes.  However, multi-object spectrographs will help quantify the contaminating population and increases in throughput provided by modern IR detectors should help significantly.  Ultimately, ground based 30m telescopes or space based spectroscopy from HST or JWST will be required to make significant progress towards spectroscopcially confirming this population of objects.

	Alternatively, a very promising means of confirming the redshift is through the [CII] 158$\mu$m line. This line is the dominant gas cooling line in the interstellar medium, tracing the cold neutral medium and photon-dominated regions at the interface with active star forming regions, and it is now being routinely detected at $z > 4$ using existing sub-mm telescopes \citep{2009A&A...500L...1M,2006ApJ...645L..97I,2005A&A...440L..51M}.  \citet{2009Natur.457..699W}  show that the [CII] line is a critical tracer of star formation in the first galaxies.  Bootstrapping from the empirical relationship between [CII] luminosity and star formation rate based on existing high-z samples, we expect a peak line flux density of $\sim 1$ to 3 mJy for these sources. Such a line could be detected with the Plateau de Bure interferometer at 240 GHz in about 10 hours  \citep{2009A&A...500L...1M}.  Furthermore, once operational in 2013, ALMA will allow for high resolution imaging of the gas on sub-kpc scales in even short integrations.
	 	
\subsection{Implications for galaxy studies of the $z>7$ universe}

\begin{figure}
\includegraphics[scale=0.34 ]{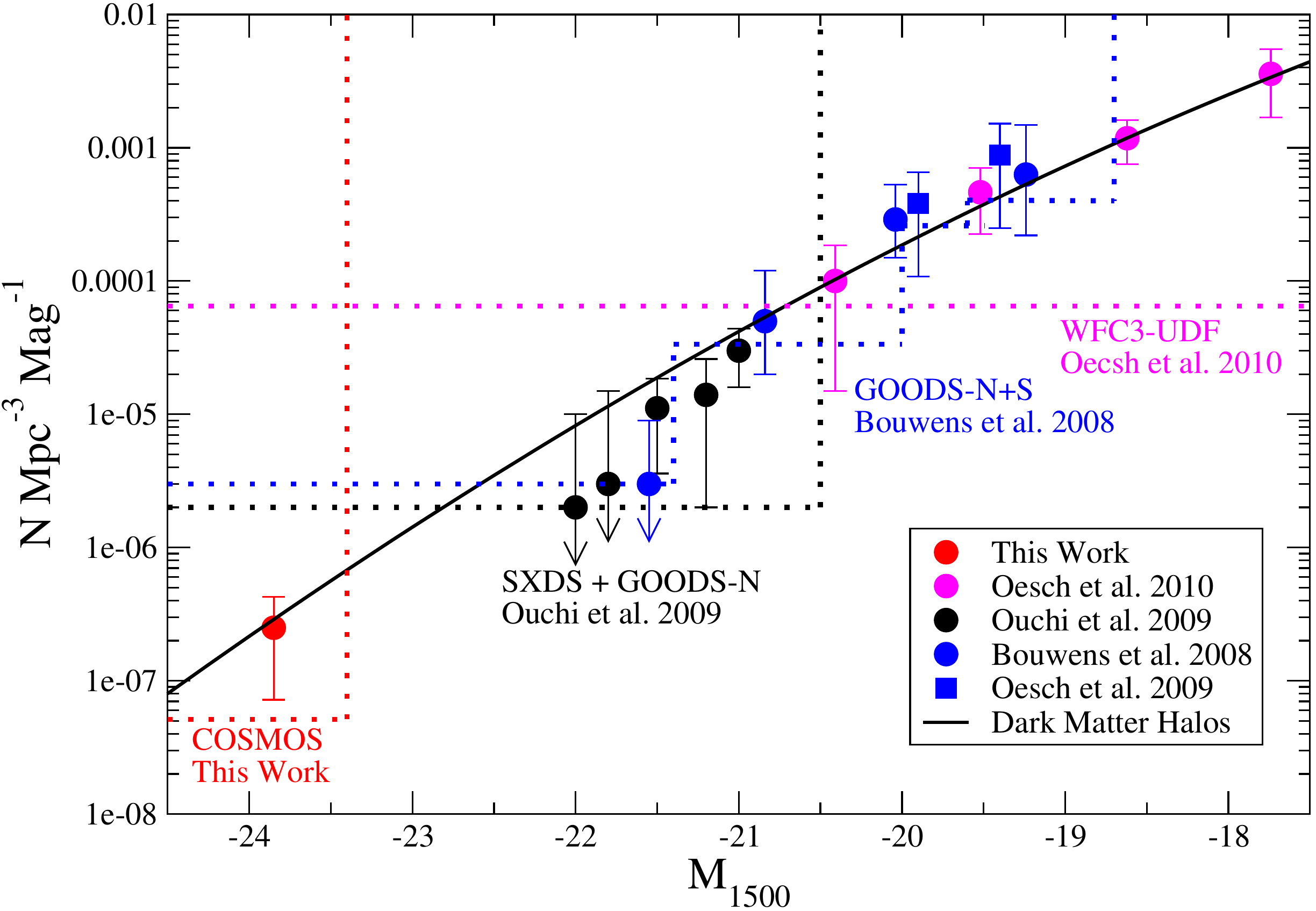}
\caption{\label{f:lf-evolution}Constraints on the $z\sim7$ galaxy luminosity function from ground and space based data  \citep{2010ApJ...709L..16O,2009ApJ...706.1136O,2009ApJ...690.1350O,2008ApJ...686..230B} are shown along with the expected distribution of dark matter haloes (black line) from \citet{2008ApJ...686..230B} re-normalized to the WFC3 points at the faint end of the luminosity function.  The dotted lines indicate the limiting magnitude and survey volume limit for the various surveys.  The error bars on upper limits indicate the point at which 3 objects would be detected in the surveyed area to give an estimate of cosmic variance.  Note the GOODS and UDF data are insufficient to probe the bright end of the luminosity function and the density of sources is consistent with the expected dark matter halo distribution.}
\end{figure}

	 A change in the shape of the luminosity function at $z>6$ is not unexpected.  In Cold Dark Matter (CDM) models the knee of the galaxy luminosity function is usually explained by feedback which truncates star formation in the highest mass haloes.   If feedback is inefficient in the first generation of galaxies, their luminosity function should flatten out to follow the power-law halo mass function at these redshifts.  The galaxy number density measured in this paper is below the expected halo density,  \citep{2006MNRAS.369.1869C,2008ApJ...686..230B,2009ApJ...706.1136O,2010ApJ...709L..16O}, so we can not rule these objects out based on space density alone.   The most likely feedback mechanisms are Active Galactic Nuclei (AGN), supernova, and stellar winds.  But, in the early universe  these mechanisms are likely to be inefficient.  AGN feedback does not become efficient until the central black hole grows to a large enough mass to provide significant energy output.  Simulation suggest this takes $\sim$200-500 Myr, leaving plenty of time for bright galaxies to exist at $z>6$  \citep{2007ApJ...665..187L}.  Furthermore, simulations have shown that supernova do not contribute enough energy to truncate star formation, and may actually enhance star formation in a gas rich environment \citep{Wise:2008p657}.  Finally, gas in the early universe will have very low metalicity, making winds less efficient at removing material.

	 Several empirical studies also support the existence of bright starbursts at $z>7$.  Studies of the $z\sim4-6$ mass function point to a epoch of star formation at $z>6$ \citep{Stark:2009p42,2009MNRAS.395.2196M}, and few studies suggest even more massive objects may exist  \citep{2008ApJ...676..781W,2007MNRAS.374..910E}, but there is significant disagreement on this latter point \citep[see ][]{2007MNRAS.376.1054D,2007ApJ...665..257C}.  Furthermore, $z\sim6$ quasars also exhibit significant enrichment \citep{2007ApJ...669...32K} and high dust contents \citep{2007ApJ...671L..13R} implying a significant earlier generation of stars was present. 
	 
	 If we take the density of M$\simeq10^{11}$M$_\sun$ galaxies to be $\sim1 \times 10^{-6}$ dex$^{-1} Mpc^{-3}$ at $z\sim6$ \citep{Stark:2009p42,2009MNRAS.395.2196M}, the continuos star formation rate required to form these objects between $z\sim20$ and $z\sim6$ is $>130$ M$_\sun$ yr$^{-1}$, which implies an intrinsic $J$ band magnitude of $J<23.6$ at $z=7$, placing an upper limit on the density of these systems of $\sim 4 \times 10^{-7}$ mag$^{-1}$ Mpc$^{-3}$.  Star formation in bursts would make these objects even more luminous, but also rarer because they are short lived.  Assuming a minimum formation time of 100 Myr for a $10^{11}$M$_\sun$ galaxy yields a lower limit on the space density of $\sim 5 \times 10^{-8}$ mag$^{-1}$ Mpc$^{-3}$.	  Our measured value is reasonably between these two extremes.  
	 	
\begin{figure}
\includegraphics[scale=0.4]{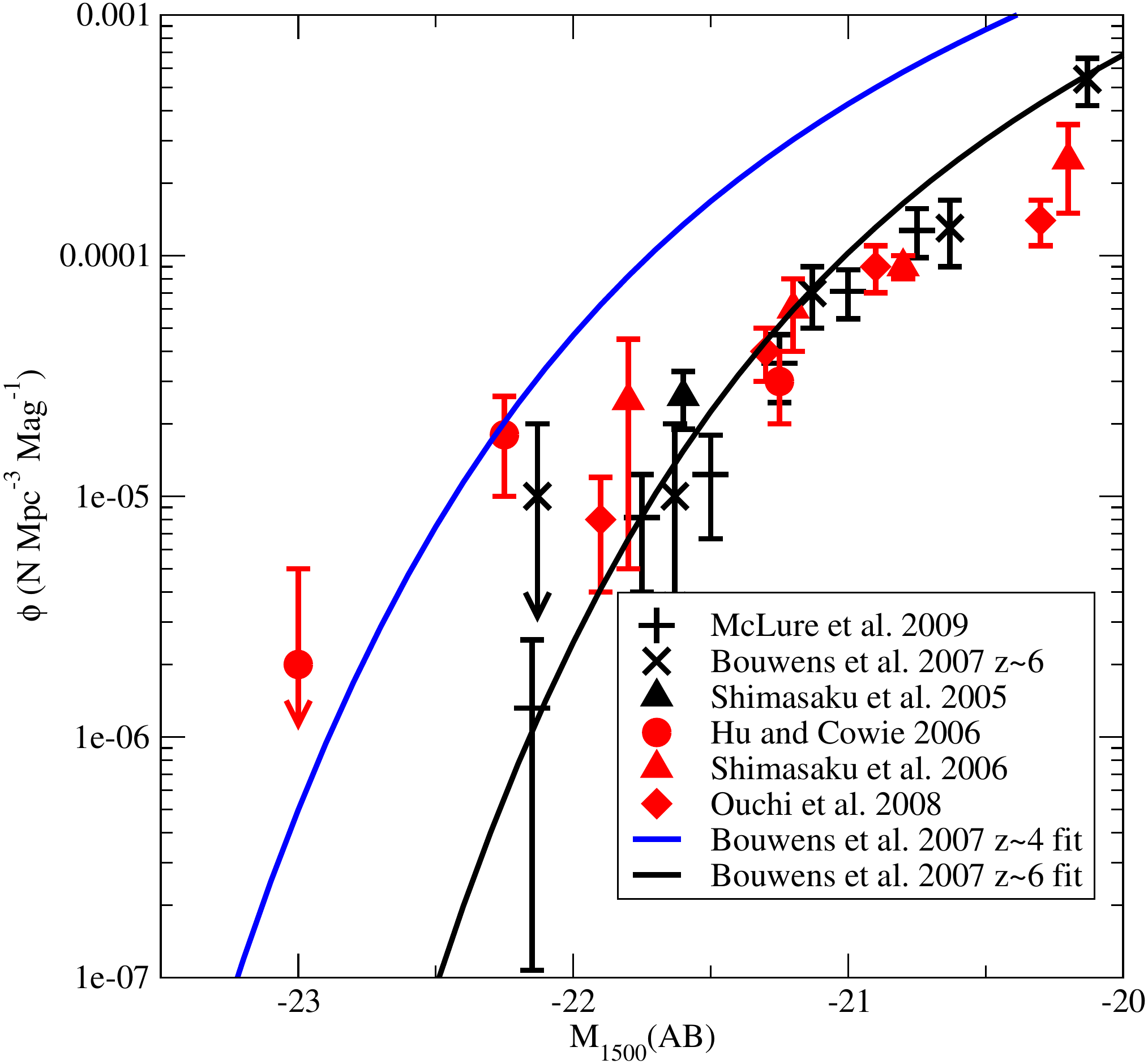}
\caption{\label{f:lae-lf} Data on the $z\sim6$ UV (1500\AA)  luminosity function from the literature is shown with LBG samples plotted in black \citep{2009MNRAS.395.2196M, 2007ApJ...670..928B,2005PASJ...57..447S} and $z\sim5.7$ LAE samples plotted in red \citep{2006Natur.440.1145H, 2006PASJ...58..313S, 2008ApJS..176..301O}.  The LAE samples are truncated at the continuum magnitude where they begin to be incomplete.  The best fit luminosity functions at $z\sim4$ and $z\sim6$ from \citet{2007ApJ...670..928B} are shown.  We note that the LBG and LAE data are statistically consistent with the exception of the \citet{2009MNRAS.395.2196M} results which are based on photometric redshifts. }

\end{figure}

	In addition, there is significant disagreement about the evolution of the bright end of the luminosity function at $z>4$, so the scale of disagreement between our result and those at $z\sim6$ is unclear.  Deep ACS surveys suggest the density of bright galaxies is dropping at $z>4$  \citep{2008ApJ...686..230B} .  However, these surveys cover a very small volume and hence lack significant constraints on the bright end of the luminosity function.  Larger area ground based surveys provide mixed conclusions with \citet{2006ApJ...648..299S} and \citet{2007MNRAS.376.1557I} pointing to no evolution in the density of bright galaxies while \citet{2006ApJ...653..988Y} and \citet{2008ASPC..399...54C} find mild evolution.  \citet{2009MNRAS.395.2196M} find strong evolution, however their results are based on photometric redshfits which can miss a significant number of bright objects at these redshifts (Capak et al. in prep).  In any case, the continuum luminosity function of spectroscopically confirmed Ly-$\alpha$ emitters shown in Figure \ref{f:lae-lf} also places a lower limit on the luminosity function at $z\sim5.7$ which is higher than the estimate from many studies at $z\sim6$ as is also noted by \citet{2006Natur.440.1145H}.  
	
	While it is numerically possible to fit a luminosity function to our data at $z\sim7$ the results would be misleading given the present uncertainty in the data.  In addition to the uncertainty in the redshift of our sources, other surveys are likely dominated by cosmic shot noise at the bright end due to their small volumes \citep{2009MNRAS.394..577O}.   At $z>5$ the luminous galaxy population is extremely clustered, with bias factors of  $\sim6-10$ \citep{2009A&A...498..725H, 2004ApJ...611..685O}.  This means the mode  (most common) luminosity function probed by small area surveys is likely to systematically underestimate the density of luminosity galaxies because the massive haloes which host these massive (bright) systems are unlikely to occur in the survey volume. To account for this uncertainty we have set the upper limits in Figure \ref{f:lf-evolution} to the point where they would detect 3 galaxies in an attempt to account for the presently unknown galaxy clustering properties.

\section{Conclusions}	

	The key to any successful Lyman break selection criteria is differentiating between the Lyman break and the 4000\AA\ break at lower redshifts and separating these objects from stars.  For a $2<z<4$ LBG selection the low redshift contamination is from $z<0.5$ where the galaxy population is well understood along with O, B, and A stars which are relatively rare \citep{2003ApJ...592..728S}.  For a $4<z<6$ selection contamination is from $0.5<z<1.5$ galaxies where the population is not as well understood and M stars which are much more common \citep{2008MNRAS.385..493S,2004AJ....127..180C}.  Furthermore, spectroscopic confirmation of both the high and low redshift sources found by a $4<z<6$ LBG selection is more difficult.  Since these selection effects are not well understood this becomes a likely source of the large discrepancies seen between $z\sim5-6$ galaxy studies \citep{2008MNRAS.385..493S,2007ApJ...670..928B, 2007MNRAS.376.1557I}.  For a $z>7$ LBG selection contamination comes from $1.5<z<2.5$ galaxies and L,T, and Y dwarfs, all of which are poorly understood and extremely difficult to spectroscopically confirm.  This is mitigated somewhat by Spitzer IRAC and $K_s$ band data which place strong constraints on the SED shape  \citep{2008MNRAS.386..370S,2006ApJ...649L..67L}.  However, the strength of these constraints is limited by the lack of a representative set of object templates at $1.5<z<2.5$ and the large uncertainties on deep Spitzer photometry due to confusion and under-sampling.
	
	Of our three candidates, we find object 1 is likely, but not conclusively at $z=7.7$,  object 2 is at $z\sim1.6$ and object 3 is very likely at $z=6.95$.  However, the faint $H$ band flux of object 3 is puzzling, suggesting a strong 2175\AA\ dust absorption feature at $z=6.95$ or a Quasar type spectra with strong Ly-$\alpha$ and CIV emission at $z=8.43$. The spectral features observed in these sources are weak but known galaxy SEDs fail to produce a $z<<7$ solution for these objects, and no objects at brighter magnitudes exhibit a similar SED, so together the evidence is reasonably strong.
		
	Whether or not these sources are at $z>7$, the difficulty in confirming even these bright objects suggests that deep small area surveys with small numbers of filters provide poor constraints on the $z>7$ galaxy population. If the presented galaxies are at $z>7$, then galaxies likely follow the dark matter halo mass function without a cut-off at the high-mass end as is observed at low redshift.  As a result pencil beam surveys will fail to provide an accurate census of the galaxy population due to a lack of volume which prevents them from finding the most luminous galaxies and cosmic variance which increases their error.  If we assume these galaxies are contaminated by $z<<7$ objects, the z-dropout selection function needs to be quantified by spectroscopy before color selections can be used.  If high equivalent width line emitters are the main source of contamination, then current NIR spectroscopy instruments may be able to determine this.  However, direct spectroscopic confirmation of the $z>7$ galaxy population will not be possible until thirty meter telescopes, JWST, or possibly ALMA is available.  Either way, medium deep, pan-chromatic observations with adjacent filters and deep $i$ and $z$ band exposures will provide the best constraints on the $z>7$ population for the foreseeable future.
	
\acknowledgements
The authors wish to recognize and acknowledge the very significant cultural role and reverence that the summit of Mauna Kea has always had within the indigenous Hawaiian community.  We are most fortunate to have the opportunity to conduct observations from this mountain.  Support for this work was provided by the Spritzer Science Center which is operated by the Jet Propulsion Laboratory (JPL), California Institute of Technology under NASA contract 1407, NASA through contract 1278386 issued by the JPL and NASA grant HST-GO-09822.  This work is based in part on observations made with the Spitzer Space Telescope, which is operated by the Jet Propulsion Laboratory, California Institute of Technology under a contract with NASA. Support for this work was provided by NASA through an award issued by JPL/Caltech.  This research has made use of the NASA/ IPAC Infrared Science Archive, which is operated by the Jet Propulsion Laboratory, California Institute of Technology, under contract with the National Aeronautics and Space Administration.  HJM and JPK acknowledge support from the French Agene National de la Recheche fund ANR-07-BLAN-0228 as well as from CNES and the Programme National Cosmologie et Galaxies.  PC would like to acknowledge the Keck remote observing staff who allowed him to simultaneously attend the NIRSPEC observations and the birth of his daughter.

\bibliography{ms}

\end{document}